\documentclass[11pt]{article}
\usepackage[a4paper,margin=2.15cm]{geometry}
\usepackage{amsmath,amssymb,bm,mathtools}
\usepackage{graphicx}
\usepackage{booktabs,array,multirow}
\usepackage[numbers,sort&compress]{natbib}
\usepackage{microtype}
\usepackage{xcolor}
\usepackage{hyperref}
\usepackage{authblk}
\usepackage{caption}
\usepackage{subcaption}
\usepackage{siunitx}
\usepackage{enumitem}
\graphicspath{{../figures/}}
\hypersetup{colorlinks=true,citecolor=blue!55!black,linkcolor=blue!55!black,urlcolor=blue!55!black}
\newcommand{\dd}{\mathrm{d}}
\newcommand{\kB}{k_{\mathrm B}}
\newcommand{\Pin}{P_{\odot,\mathrm{inc}}}
\newcommand{\etal}{\textit{et al.}}

\title{Fundamental limits of hot-carrier and photothermal plasmonic solar chemistry}
\author[1,2,*]{Seungwoo Lee}
\affil[1]{KU-KIST Graduate School of Converging Science and Technology, Korea University, Seoul 02841, Republic of Korea}
\affil[2]{Department of Integrative Energy Engineering (College of Engineering) and Department of Biomicrosystem Technology, Korea University, Seoul 02841, Republic of Korea}
\affil[*]{Email: \href{mailto:seungwoo@korea.ac.kr}{seungwoo@korea.ac.kr}.}
\date{}

\begin{document}
\maketitle

\begin{abstract}
Plasmonic catalysts convert sunlight through direct metal--adsorbate excitation, nonequilibrium hot carriers and photothermal heating, yet these channels lack a common efficiency limit analogous to the Shockley--Queisser (SQ) limit. Here, we formulate a channel-resolved detailed-balance theory for plasmonic solar chemistry. The theory enforces exclusive partition of absorbed power, microscopic reversibility, non-negative entropy production, electromagnetic passivity and causality, and one physical nanostructure shared by all wavelengths and operating states. A reduced tightly coupled cycle obeys a chemical diode law with closed-form stall and maximum-power free energies. It recovers a 33.68\% single-threshold radiative limit and the 30.58\% ideal single-junction water-splitting limit at a 1.23-eV chemical load, whereas a Fowler hot-carrier kernel lowers the limit to 8.53\%. Finite one-structure benchmarks show that independent wavelength optimization overestimates shared-geometry performance by 6.6--33.8\%, while cross-frequency correlations tighten semidefinite upper bounds by 21--56\%. An exact multi-electron network yields a universal arrival--storage criterion and the low-flux scaling $J_n\sim g^n\tau^{n-1}$. A source-conditioned gold/p-type gallium nitride case study and a bias- and separation-complete ledger then distinguish enhanced product formation from net solar-energy conversion. The resulting framework provides a thermodynamically closed benchmark for hot-carrier and photothermal plasmonic chemistry.
\end{abstract}

\section*{Introduction}
Detailed-balance limits provide a reference against which energy-conversion technologies can be evaluated before particular materials or device architectures are optimized. The Shockley--Queisser (SQ) theory achieves this for a single-junction solar cell by reducing the converter to its absorption threshold, radiative emission and operating temperature\cite{ShockleyQueisser1961}. Extensions have incorporated non-ideal spectra, multijunctions, hot carriers, carrier multiplication, optical concentration and nanophotonic light management\cite{Henry1980,RossNozik1982,HannaNozik2006,YuFan2010,Rau2014,Polman2016}. In photoelectrochemistry, photodiode detailed balance has been combined with catalytic overpotentials and transport losses to create a unified limit for photoelectrochemical (PEC) water splitting\cite{Bolton1985,Fountaine2016}. General photochemical thermodynamics predates both developments: (i) Ross related incident radiation to an attainable photochemical potential\cite{Ross1967}, and (ii) subsequent work derived yield limits for one- and multichromophore systems\cite{RossHsiao1977,Fingerhut2010}. Detailed balance has also been applied to ideal semiconductor hot-carrier photocatalysts\cite{Takeda2022}, while thermodynamic models of the plasmoelectric effect describe light-induced electrochemical potentials without a reaction-resolved solar-to-chemical ledger\cite{Sheldon2014,Sheldon2016}. These theories demonstrate that a chemical product can be treated as a thermodynamic load, but they do not resolve the competing microscopic channels through which a plasmonic metal transfers optical energy to a reaction.

Plasmonic solar chemistry is intrinsically multichannel. Localized surface plasmons can drive direct transitions between hybridized metal--adsorbate states, generate nonthermal electron--hole distributions that subsequently reach an interface, or relax into lattice heat that accelerates thermal chemistry\cite{Linic2011,Christopher2011,Christopher2012,Mukherjee2013,Boerigter2016,Brongersma2015}. Hybrid antenna--reactor structures add spatial transfer and material-selective dissipation, while interband and intraband excitations can favor different products\cite{Sundararaman2014,Brown2016,Cortes2017,Kiani2024,Kiani2026}. Because unextracted carrier energy ultimately becomes heat, hot-carrier, and photothermal contributions cannot be added as independent gains. Nor can extinction be identified with useful absorption: scattering, parasitic electronic loss, carrier thermalization, interface transmission, and chemical back reactions each intervene between the incident photon and stored chemical free energy\cite{Leenheer2014,ZhangYamSchatz2016,DubiSivan2019,Sivan2020}.

A second difficulty is electromagnetic realizability. Large local fields or spectrally optimized absorption curves do not by themselves define a physical broadband catalyst. Passivity, causality, and material sum rules constrain the integrated response of every scatterer\cite{Miller2014,Miller2016,ZhangMonticoneMiller2023}. Moreover, responses optimized separately at different frequencies or operating states may correspond to different geometries. Cross-scenario correlations are therefore needed to require one common nanostructure\cite{Shim2024}. This distinction is especially consequential under sunlight, where a converter must simultaneously process a broad spectrum, two polarizations, a range of incidence directions, and reaction-dependent optical states.

\begin{figure}[t]
\centering
\includegraphics[width=\textwidth]{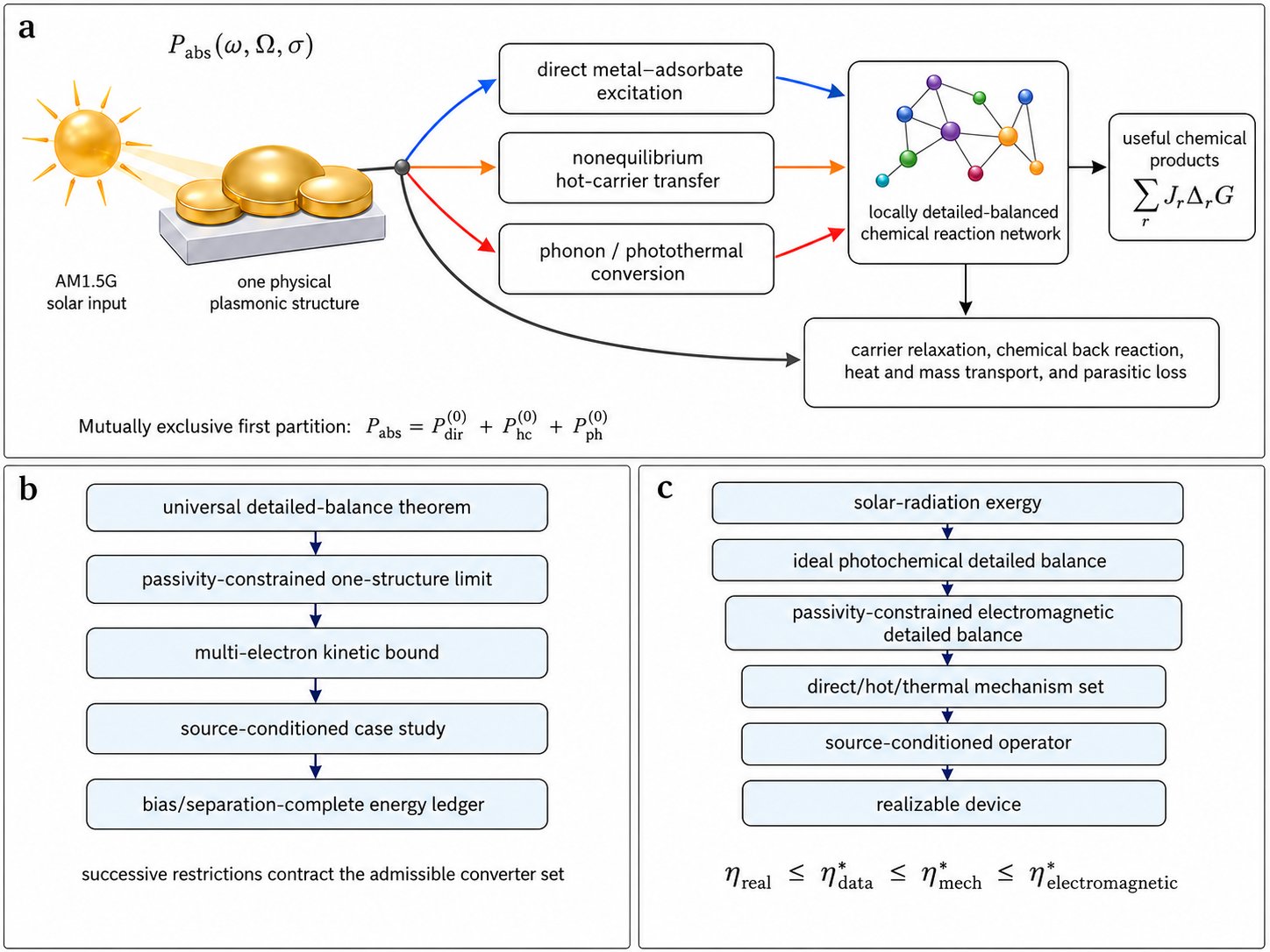}
\caption{\textbf{Unified framework, logical sequence and hierarchy of limits.} \textbf{a}, Air Mass 1.5 Global (AM1.5G) solar radiation illuminates one physical plasmonic structure. For a mode specified by angular frequency $\omega$, propagation direction $\Omega$, and polarization $\sigma$, the absorbed power $P_{\rm abs}(\omega,\Omega,\sigma)$ is assigned exactly once to direct metal--adsorbate excitation, nonequilibrium hot-carrier generation, or initially thermalized phonon/photothermal power. Blue and orange arrows identify quantum channels that can enter a chemical reaction network constrained by local detailed balance, the red arrow identifies thermal input, and grey arrows collect carrier relaxation, chemical back reaction, heat and mass transport, and parasitic loss. Only product flux weighted by stored Gibbs free energy enters the useful-output box. \textbf{b}, The five results developed in the manuscript, in their order of use: the universal detailed-balance theorem defines the thermodynamic feasible set; passivity, causality, and one-common-structure constraints bound the electromagnetic response; multi-electron accumulation bounds productive flux; source observations contract the admissible channel operators; and the final system ledger subtracts externally supplied electrical work and minimum product-separation work. \textbf{c}, Nested upper-bound hierarchy. Solar-radiation exergy bounds ideal photochemical detailed balance, which bounds passivity-constrained electromagnetic detailed balance, mechanism-constrained direct/hot-carrier/thermal conversion, a source-conditioned operator and, finally, a realizable device. Each added constraint contracts the feasible set, so the corresponding efficiency cannot increase.}
\label{fig:framework}
\end{figure}

Here, we establish a theory-first framework for the fundamental limits of hot-carrier and photothermal plasmonic solar chemistry. Figure~\ref{fig:framework}a follows the energy of one absorbed photon through mutually exclusive direct-interfacial, nonequilibrium-carrier, and phonon/photothermal branches, and then through useful chemistry or independent relaxation, reverse-reaction, transport, and parasitic-loss channels. Figure~\ref{fig:framework}b states the five results developed in sequence: a universal detailed-balance theorem, a passivity-constrained one-structure limit, an exact multi-electron kinetic bound, a source-conditioned case study, and a bias- and separation-complete energy ledger. Figure~\ref{fig:framework}c places these results in a nested hierarchy from solar-radiation exergy to a realizable device. The theory does not require state-resolved optical transition matrices or a complete transition-state landscape. Instead, it defines nested admissible channel-operator and kinetic sets that microscopic data may progressively narrow. Supplementary Note~1 formalizes these admissible sets, while Supplementary Table~1 provides the notation and claim-boundary glossary used throughout the Article.

\section*{Results}
\subsection*{A universal detailed-balance theorem}
We describe the incident radiation by modes $m=(\omega,\Omega,\sigma)$, denoting angular frequency, propagation direction, and polarization. Unless stated otherwise, every optical or system quantity denoted by $P$ is an areal power density in W~m$^{-2}$, and every free energy in an exponential is expressed per event in joules; values reported in electronvolts are converted using the elementary charge $q$. For each mode, $P_{\rm ext}$, $P_{\rm sca}$ and $P_{\rm abs}$ denote extinction, scattering, and absorption power, respectively, and obey $P_{\rm ext}=P_{\rm sca}+P_{\rm abs}$. Only absorption can initiate local conversion, and its first irreversible branching is defined as
\begin{equation}
P_{\rm abs}(m)=P_{\rm dir}^{(0)}(m)+P_{\rm hc}^{(0)}(m)+P_{\rm ph}^{(0)}(m).
\label{eq:partition}
\end{equation}
Here, $P_{\rm dir}^{(0)}$, $P_{\rm hc}^{(0)}$ and $P_{\rm ph}^{(0)}$ are the direct metal--adsorbate, nonequilibrium hot-carrier, and initially thermalized powers, respectively; the superscript $(0)$ denotes this initial partition before subsequent carrier relaxation or chemical conversion. Energy in the first two branches that fails to complete a chemical step subsequently enters heat. Equation~\ref{eq:partition} therefore excludes the common double counting, in which the same absorbed photon is credited once as a hot carrier and again as photothermal input.

Catalyst, adsorbate, and charge-storage configurations are represented by states $i$ and $j$. An elementary transition $i\rightarrow j$ coupled to reservoir or channel $a$ has probability current
\begin{equation}
J_{ij}^{(a)}=k_{ij}^{(a)}p_i-k_{ji}^{(a)}p_j,
\label{eq:edgecurrent}
\end{equation}
where $p_i$ and $p_j$ are state probabilities and $k_{ij}^{(a)}$ and $k_{ji}^{(a)}$ are the corresponding forward and reverse rate constants. Thus, $J_{ij}^{(a)}$ has units of inverse time per normalized catalyst site and is positive in the $i\rightarrow j$ direction. Steady-state probability conservation requires $\sum_{j,a}J_{ji}^{(a)}=0$ for every state $i$. Thermal and electrochemical steps satisfy local detailed balance,
\begin{equation}
\ln\frac{k_{ij}^{(a)}}{k_{ji}^{(a)}}=-\frac{\Delta E_{ij}-\sum_{\ell}\mu_{\ell}\Delta N_{\ell,ij}}{\kB T_a},
\label{eq:ldb}
\end{equation}
Here, $\Delta E_{ij}=E_j-E_i$ is the internal-energy change of the system, $\Delta N_{\ell,ij}=N_{\ell,j}-N_{\ell,i}$ is the change in particle number of chemical species $\ell$, $\mu_{\ell}$ is its reservoir chemical potential, $T_a$ is the temperature of reservoir $a$, and $k_{\mathrm B}$ is Boltzmann's constant. A nonthermal carrier reservoir is instead evaluated from its energy-resolved distribution, Pauli factors and transition kernel rather than being assigned an artificial electronic temperature. The entropy-production rate is
\begin{equation}
\dot S_{\rm gen}=\kB\sum_{i<j,a}J_{ij}^{(a)}
\ln\!\left(\frac{k_{ij}^{(a)}p_i}{k_{ji}^{(a)}p_j}\right)\geq0.
\label{eq:entropy}
\end{equation}
Here, $\dot S_{\rm gen}$ is the total entropy-production rate, the restriction $i<j$ counts each bidirectional edge once, and the sum over $a$ includes every optical, thermal, chemical and electrical reservoir coupled to that edge. This expression places optical excitation, carrier relaxation, chemical back reaction, and heat flow in one thermodynamic network and follows the non-negative entropy-production structure of stochastic chemical thermodynamics\cite{Seifert2012,RaoEsposito2016}. Supplementary Note~2 proves Eq.~\ref{eq:entropy} edge by edge and shows how state, site, and atomic conservation enter the global feasible set.

For useful reactions $r$ that store positive Gibbs free energy $\Delta_rG$ at the ambient state, the net solar-to-chemical efficiency is bounded by the plasmonic solar-chemical detailed-balance (PSCD) variational theorem
\begin{equation}
\boxed{\displaystyle
\eta_{\rm PSCD}^{\star}=\frac{1}{\Pin}
\sup_{\mathcal F}\left[
\sum_r \mathcal J_r\Delta_rG-P_{\rm bias}-P_{\rm sep}
\right].}
\label{eq:master}
\end{equation}
In Eq.~\ref{eq:master}, $\mathcal J_r$ is the net areal molar flux of useful reaction $r$, $\Delta_rG>0$ is the molar Gibbs free energy stored by that reaction at the declared reference state, $\Pin$ is the incident solar power density, and the superscript $\star$ denotes the supremum over all admissible converters. The feasible set $\mathcal F$ contains Maxwell's equations, passivity, causality, electromagnetic sum rules, one common geometry for all scenarios, Eq.~\ref{eq:partition}, state and atomic conservation, Eq.~\ref{eq:ldb}, heat and mass transfer, and Eq.~\ref{eq:entropy}. $P_{\rm bias}$ is externally supplied electrical power density, and $P_{\rm sep}$ is the minimum reversible product-separation power density. The theorem deliberately evaluates stored Gibbs free energy, rather than activation energy or the enthalpy already present in exergonic reactants. Consequently, light-accelerated oxidation can exhibit a large rate enhancement, while storing little or no solar free energy.

The admissible channel set can be nested according to available information. A universal positive-semidefinite partition defines $\mathcal C_{\rm univ}$; carrier thresholds, survival, and transfer restrictions define $\mathcal C_{\rm mech}\subset\mathcal C_{\rm univ}$; and wavelength-resolved observables define $\mathcal C_{\rm data}\subset\mathcal C_{\rm mech}$. The resulting hierarchy is
\begin{equation}
\eta_{\rm data}^{\star}\leq\eta_{\rm mech}^{\star}\leq\eta_{\rm electromagnetic}^{\star}\leq\eta_{\rm ideal\,photochemical}^{\star}.
\label{eq:nested}
\end{equation}
Here, $\eta_{\rm data}^{\star}$ uses source-conditioned operators, $\eta_{\rm mech}^{\star}$ additionally enforces carrier-generation and transfer mechanisms, $\eta_{\rm electromagnetic}^{\star}$ retains only the passive-causal electromagnetic constraints, and $\eta_{\rm ideal\,photochemical}^{\star}$ is the ideal photochemical detailed-balance limit. Raw electronic-structure transition matrices would further contract $\mathcal C_{\rm data}$, but their absence does not invalidate the broader inequalities. This hierarchy therefore preserves a strict distinction between universal, mechanism-constrained, and source-conditioned claims.

\subsection*{Chemical diode law and reduction to established limits}
The universal network admits an analytic reduction, when a single tightly coupled cycle stores chemical free energy $\mu$ per completed event. Let $J(\mu)$ be the net cycle-completion flux, $G$ the useful excess photogeneration flux at zero chemical load, $R_0$ the equilibrium reverse-cycle prefactor, and $T_0$ the ambient temperature. Microscopic reversibility gives the chemical analogue of the illuminated diode law,
\begin{equation}
J(\mu)=G-R_0\left[\exp\!\left(\frac{\mu}{\kB T_0}\right)-1\right].
\label{eq:chemicaldiode}
\end{equation}
The stall, or chemical open-circuit, free energy is
\begin{equation}
\mu_{\rm oc}=\kB T_0\ln\!\left(1+\frac{G}{R_0}\right).
\label{eq:muoc}
\end{equation}
Maximizing $P(\mu)=\mu J(\mu)$ yields a closed-form maximum-power load. With $\gamma=G/R_0$ and $\xi=\mu_{\rm mp}/(\kB T_0)$,
\begin{equation}
\exp(\xi)(1+\xi)=1+\gamma,\qquad
\mu_{\rm mp}=\kB T_0\left\{W[e(1+\gamma)]-1\right\},
\label{eq:lambert}
\end{equation}
where $W$ is the principal branch of the Lambert $W$ function, $e$ is Euler's number, $\xi=\mu_{\rm mp}/(\kB T_0)$ is the dimensionless maximum-power load, and $\gamma=G/R_0$ is the generation-to-reverse-rate ratio. The maximum chemical power density is $P_{\rm max}=\kB T_0R_0\xi^2e^\xi$. Supplementary Note~3 gives the full derivation of this maximum-power solution, and Supplementary Figure~1a shows how the resulting load remains below the reversible stall free energy over the threshold range.

\begin{figure}[t]
\centering
\includegraphics[width=\textwidth]{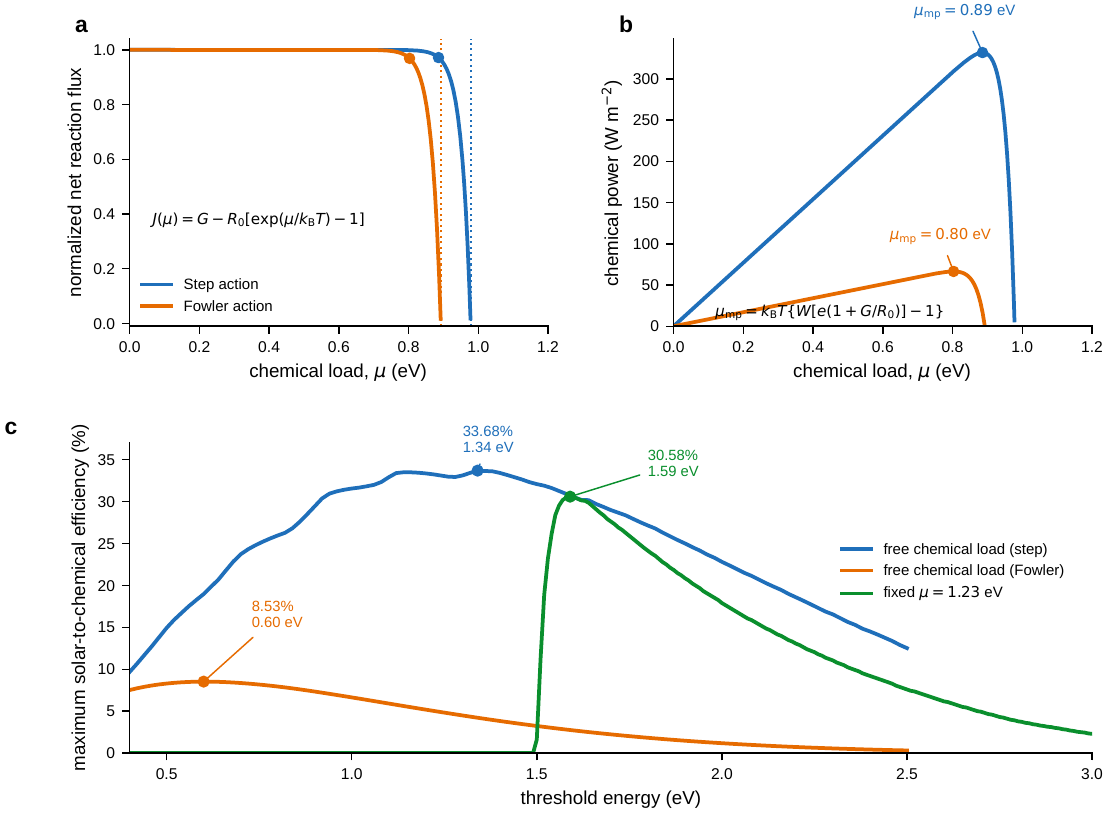}
\caption{\textbf{Chemical diode theorem and reduction checks.} \textbf{a}, Normalized net reaction flux $J(\mu)$ as a function of chemical load $\mu$ for a unit-step action spectrum (blue) and a Fowler hot-carrier action (orange). The nearly horizontal low-load regime is limited by useful photogeneration, whereas the abrupt high-load decrease is produced by the reciprocal reverse-cycle term in Eq.~\ref{eq:chemicaldiode}. Dotted vertical lines mark the chemical open-circuit, or stall, free energies from Eq.~\ref{eq:muoc}; filled circles mark the maximum-power loads. \textbf{b}, Chemical power $P(\mu)=\mu J(\mu)$ for the same two actions. The labelled maxima, $\mu_{\rm mp}=0.89$ and 0.80~eV for the displayed examples, coincide with the closed-form Lambert-$W$ solution in Eq.~\ref{eq:lambert}. \textbf{c}, Maximum solar-to-chemical efficiency versus threshold energy under the ASTM International G173 global-tilt reference spectrum. The blue variable-load unit-step curve reproduces the Shockley--Queisser (SQ) envelope; the green curve fixes $\mu=1.23$~eV and reproduces the ideal single-junction photoelectrochemical (PEC) water-splitting limit; and the orange Fowler-filtered curve quantifies the loss associated with a restricted population of extractable high-energy carriers. Markers and labels give the maximum efficiency and optimum threshold of each curve.}
\label{fig:diode}
\end{figure}

Figure~\ref{fig:diode}a plots normalized net reaction flux versus chemical load for a unit-step action and an aggregated Fowler hot-carrier action. The solid curves remain generation-limited at small $\mu$, then collapse as the reciprocal reverse current rises; dotted vertical lines mark the stall loads and filled circles mark the maximum-power loads. Figure~\ref{fig:diode}b converts the same currents to chemical power and shows that the analytic Lambert-$W$ solution locates the maxima without numerical fitting; the associated chemical fill factors are resolved in Supplementary Figure~1b. Figure~\ref{fig:diode}c evaluates the maximum efficiency over threshold energy using the ASTM International G173 (ASTM G173) global-tilt reference spectrum, whose integrated irradiance is \SI{1000.3707}{W.m^{-2}}. A freely optimized unit-step converter reaches 33.68\% at 1.34~eV, reproducing the terrestrial single-junction SQ envelope. Fixing the chemical load at 1.23~eV gives 30.58\% at 1.59~eV, matching the ideal single-junction PEC water-splitting limit\cite{Fountaine2016}. The Fowler action
\begin{equation}
Y_{\rm F}(E;E_{\rm th})=\left(\frac{E-E_{\rm th}}{E}\right)^2\Theta(E-E_{\rm th}),
\label{eq:fowler}
\end{equation}
where $Y_{\rm F}$ is the dimensionless Fowler extraction action, $E$ is photon energy, $E_{\rm th}$ is the carrier-extraction threshold, and $\Theta$ is the Heaviside step function. This action lowers the maximum to 8.53\% at $E_{\rm th}=0.60$~eV. Together, Figure~\ref{fig:diode}a--c provide three reduction checks and separate the thermodynamic effect of a fixed chemical load from the microscopic scarcity of extractable high-energy carriers. The three limiting reductions and their numerical optima are collected in Supplementary Table~2.

\subsection*{Passivity-constrained one-structure limit}
Next, we constrain $G$ electromagnetically. An optical scenario $s$ specifies wavelength or frequency, polarization, incidence direction, and operating state. In a finite basis, $\mathbf e_s(\omega)$ is the incident-field coefficient vector, $\mathbf p_s(\omega)=\mathsf T(\omega)\mathbf e_s(\omega)$ is the induced polarization-current vector, and $\mathsf T(\omega)$ is the passive-causal scattering response operator. With the incident fields normalized to the solar spectral irradiance, a channel-resolved absorbed-power spectral density can be written
\begin{equation}
P_c^{(s)}(\omega)=\frac{\omega}{2}\mathbf p_s^{\dagger}\mathsf W_c(\omega)\mathbf p_s,
\qquad \mathsf W_c\succeq0,
\label{eq:channelpower}
\end{equation}
where $P_c^{(s)}(\omega)$ has units W~m$^{-2}$ per unit angular frequency, $c$ indexes a dissipative channel, $\dagger$ denotes Hermitian conjugation, $\mathsf W_c(\omega)$ is its Hermitian positive-semidefinite work operator, and $\succeq0$ denotes the Loewner positive-semidefinite order. The identity $\sum_c\mathsf W_c=\mathsf W_{\rm abs}$ states that the channel operators exhaust total absorption. The channel labels may distinguish direct adsorbate excitation, intraband and interband hot carriers, and initially thermalized loss. The matrix-oscillator representation of every passive scattering body expresses $\mathsf T$ through positive-semidefinite (PSD) spectral weights that obey high- and low-frequency sum rules\cite{ZhangMonticoneMiller2023}. In a finite basis these become
\begin{equation}
\mathsf X(\omega)\succeq0,
\quad \int_0^\infty\mathsf X(\omega)\dd\omega\preceq\mathsf B_{\rm H},
\quad \int_0^\infty\frac{\mathsf X(\omega)}{\omega^2}\dd\omega\preceq\mathsf B_{\rm L}.
\label{eq:sumrules}
\end{equation}
Here, $\mathsf X(\omega)$ is the positive-semidefinite matrix-valued oscillator spectral measure, $\mathsf B_{\rm H}$ and $\mathsf B_{\rm L}$ are the finite high- and low-frequency oscillator budgets, and $\preceq$ denotes the Loewner order. These inequalities prevent independent assignment of arbitrarily large resonances across the spectrum. Supplementary Note~4 derives the matrix-oscillator and channel-operator construction, and Supplementary Table~3 lists the finite response budgets used below. Its material specialization is tested in Supplementary Figure~2, with the corresponding silver (Ag), gold (Au) and rhodium (Rh) values reported in Supplementary Table~4.

Figure~\ref{fig:passivity}a shows how progressively stronger structural restrictions contract a conditional two-component matrix-oscillator problem. Frequency-dependent PSD atoms give 4.310\% hybrid efficiency, one common eigenbasis gives 4.282\%, and one polarization axis gives 2.579\%. In the outer model, \SI{225.68}{W.m^{-2}} is absorbed and partitioned into \SI{16.78}{W.m^{-2}} of direct work, \SI{25.25}{W.m^{-2}} of hot-carrier work, and \SI{1.08}{W.m^{-2}} of recoverable thermal work. These are conditional limits for stated oscillator budgets, not material-independent efficiencies.

Figure~\ref{fig:passivity}b isolates the requirement that all scenarios share one geometry in an exactly enumerable seven-voxel, three-frequency benchmark. Allowing a different structure at each frequency gives an objective of 2.535. A semidefinite-programming (SDP) relaxation without cross-frequency correlations gives 2.825; adding all cross correlations lowers the certified bound to 2.231; and the exact shared structure gives 1.678. The exact one-structure penalty is 33.8\%, while the 21.0\% reduction from the no-cross to the all-cross bound quantifies how much artificial scenario freedom is removed. Supplementary Note~5 derives the cross-scenario identities and lifted relaxation; Supplementary Table~5, and Supplementary Figure~3a give the numerical hierarchy for this benchmark.

Figure~\ref{fig:passivity}c repeats the hierarchy in a finite full-vector benchmark using measured optical constants for Au on p-type gallium nitride (Au/p-GaN)\cite{JohnsonChristy1972,Kawashima1997}, a planar reflected Green tensor, five spectral bins, two polarizations, and all $2^9$ binary masks. Independent scenario-specific masks give \SI{10.83}{W.m^{-2}} of net carbon-monoxide work, the no-cross SDP bound is \SI{24.89}{W.m^{-2}}, the all-cross bound is \SI{10.98}{W.m^{-2}}, and the exact shared mask gives \SI{8.97}{W.m^{-2}}. Thus, the shared structure incurs a 17.2\% penalty, whereas cross correlations tighten the certified outer bound by 55.9\%. Supplementary Figure~3b presents this measured-material hierarchy; Figure 3c and 3d record the dual-search convergence and numerical cross-identity certificates, respectively. The five-to-ten-bin reevaluation in Supplementary Figure~4a verifies that the shared-mask result is not a coarse spectral-quadrature artifact.

Figure~\ref{fig:passivity}d contrasts material response with broadband realizability in a quasistatic screen. At a 1-eV carrier threshold and 5-nm equivalent metal thickness, exact material limits are 6.37\%, 4.23\%, and 5.62\% for Ag, Au, and Rh, respectively. Supplementary Figure~2a quantifies the finite-loss correction to the low-loss asymptote. One fixed spheroid retains only 33\% of the Ag limit, 54\% of the Au limit and 92\% of the Rh limit; the sphere-to-fixed-shape-to-wavelength-wise hierarchy, optical-constant sensitivity and optimum fixed aspect ratios are resolved in Supplementary Figure~2b--d. Thereby, the material with the largest local response need not be the material whose response is most realizable by one broadband geometry.

\begin{figure}[t]
\centering
\includegraphics[width=\textwidth]{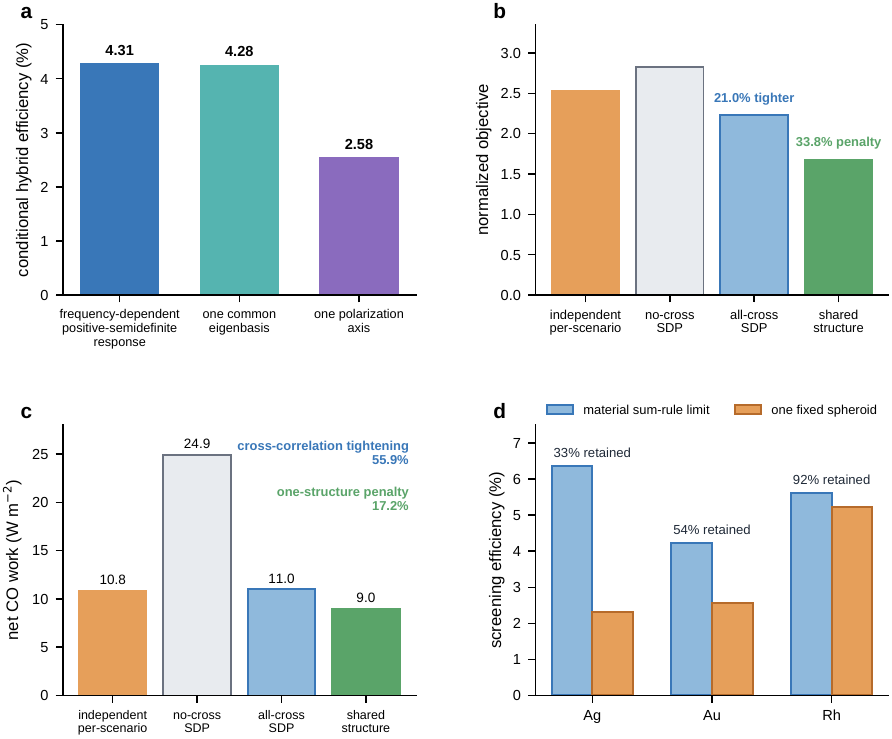}
\caption{\textbf{Passivity, causality and one-common-structure realizability.} \textbf{a}, Conditional matrix-oscillator hybrid efficiency under three nested response spaces: a frequency-dependent positive-semidefinite (PSD) response, one common two-axis eigenbasis shared by all frequencies, and one shared polarization axis. The small decrease from 4.31\% to 4.28\% shows that eigenbasis sharing is weak for the stated conditional budgets, whereas the one-axis restriction lowers the limit to 2.58\%. \textbf{b}, Exactly enumerable seven-voxel, three-frequency benchmark. Bars show the exact diagnostic obtained with independently selected structures, the semidefinite-programming (SDP) upper bound without cross-frequency identities, the all-cross SDP upper bound, and the exact shared structure. Cross-scenario identities tighten the outer relaxation by 21.0\%, and requiring one exact structure reduces the physical objective by 33.8\%. \textbf{c}, Measured-material full-vector benchmark for one gold on p-type gallium nitride (Au/p-GaN) mask. The bars report independently selected masks, the no-cross, and all-cross SDP bounds, and the exact shared mask in units of net carbon-monoxide work. Cross correlations tighten the certified bound by 55.9\%, while the common-mask requirement produces a 17.2\% exact penalty. \textbf{d}, Optimistic quasistatic material screen for silver (Ag), gold (Au) and rhodium (Rh). Blue bars are exact finite-loss material sum-rule limits, orange bars are the best one-fixed-spheroid responses, and labels give the useful-generation fraction retained by one broadband geometry. The comparison shows that a large material limit need does not imply high one-geometry realizability.}
\label{fig:passivity}
\end{figure}

\subsection*{A universal multi-electron kinetic bound}
Solar-fuel products typically require several sequential charge-transfer events. Even when the single-carrier transfer probability is high, an intermediate can decay before the next carrier arrives. Consider a productive site that must accumulate $n$ carriers. Productive arrivals occur at rate $g$, and every incomplete intermediate resets at rate $k=1/\tau$. For states $i=0,\ldots,n-1$, the stationary distribution is geometric with
\begin{equation}
\zeta=\frac{g}{g+k}=\frac{g\tau}{1+g\tau},\qquad
p_i=\frac{(1-\zeta)\zeta^i}{1-\zeta^n},
\label{eq:progression}
\end{equation}
where $k=1/\tau$ is the reset rate, $\tau$ is the lifetime of an incomplete intermediate, $p_i$ is the stationary probability of retaining $i$ carriers, and $\zeta$ is the dimensionless probability ratio between successive accumulation states. The exact product-completion rate per accumulation reservoir, $J_n$, and the productive fraction of arriving carriers, $U_n$, are
\begin{equation}
J_n=g\frac{(1-\zeta)\zeta^{n-1}}{1-\zeta^n},
\qquad
U_n=\frac{nJ_n}{g}=n\frac{(1-\zeta)\zeta^{n-1}}{1-\zeta^n}.
\label{eq:multielectron}
\end{equation}
Supplementary Note~6 derives Eqs.~\ref{eq:progression} and~\ref{eq:multielectron} directly from the stationary master equation. Their agreement with an independently assembled continuous-time Markov generator is shown in Supplementary Figure~5a,b, with a maximum relative error of $8.1\times10^{-11}$.

\begin{figure}[t]
\centering
\includegraphics[width=\textwidth]{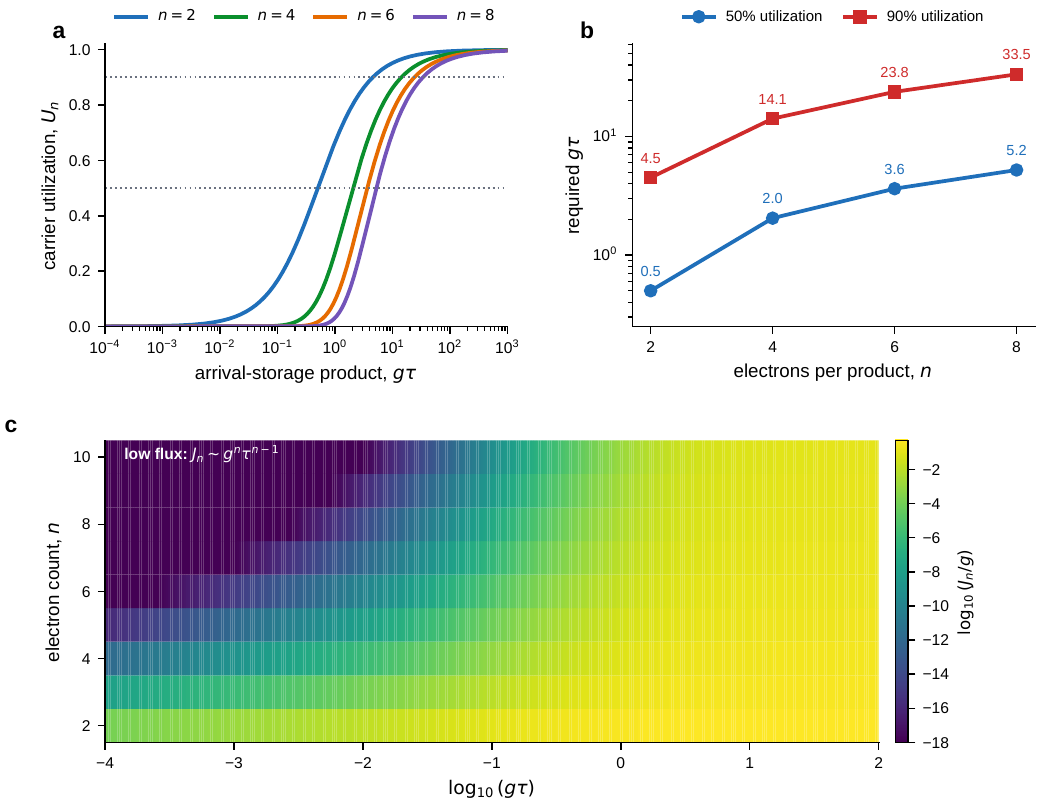}
\caption{\textbf{Universal multi-electron kinetic bound.} \textbf{a}, Exact productive-carrier utilization $U_n$ versus the dimensionless arrival--storage product $g\tau$, where $g$ is the productive carrier-arrival rate at one accumulation reservoir, $\tau$ is the lifetime of an incomplete intermediate, and $n$ is the number of carriers required per product. Curves for $n=2$, 4, 6, and 8 are compared with dotted 50\% and 90\% utilization levels. \textbf{b}, Exact values of $g\tau$ required to reach the two utilization targets in panel \textbf{a}. Numeric labels give the roots of Eq.~\ref{eq:multielectron} and show the rapidly increasing storage requirement for larger $n$. \textbf{c}, Product flux normalized by carrier-arrival flux, $J_n/g$, over electron count and $g\tau$. The colour scale is $\log_{10}(J_n/g)$; the low-flux region visualizes the asymptotic law $J_n\sim g^n\tau^{n-1}$, which suppresses high-electron-count products under dilute one-sun arrivals.}
\label{fig:multi}
\end{figure}

Figure~\ref{fig:multi}a plots the exact utilization curves for products requiring two, four, six, and eight carriers. Each curve is controlled only by the dimensionless arrival--storage product $g\tau$ and shifts to larger values as the electron count rises. Figure~\ref{fig:multi}b extracts the accumulation requirement: 50\% utilization requires $g\tau=0.50$, 2.05, 3.63 and 5.22 for $n=2$, 4, 6 and 8, while 90\% utilization requires 4.50, 14.15, 23.80 and 33.46; Supplementary Table~6 reports the exact roots used for both utilization targets. Figure~\ref{fig:multi}c displays the normalized product flux over the broader $(n,g\tau)$ plane. In the one-sun, low-flux regime $g\tau\ll1$ (equivalently $\zeta\simeq g\tau$),
\begin{equation}
J_n\sim g^n\tau^{n-1},
\label{eq:lowflux}
\end{equation}
which produces the rapidly darkening low-flux region at large $n$. The low-flux expansion and its domain of validity are given in Supplementary Note~6. The exact bound suggests four independent design levers: increase the productive arrival rate, extend the intermediate lifetime, reduce dilution over inactive reservoirs, or reduce and stage the number of simultaneously accumulated redox equivalents.

\subsection*{Source-conditioned gold on p-type gallium nitride case study}
To demonstrate how public observables narrow the universal channel set without claiming first-principles closure, we construct a source-conditioned operator for wavelength-dependent carbon-dioxide (CO$_2$) reduction on a Au/p-GaN photocathode. Operando measurements report product partitions under constant absorbed optical power, with carbon monoxide (CO) favored under 460--560-nm wavelength excitation and molecular hydrogen (H$_2$) favored at 640--800~nm wavelength\cite{Kiani2026}; related work distinguishes low-energy inner-sphere and higher-energy outer-sphere transfer in the same platform\cite{Kiani2024}. We digitize the reported CO, formate, and H$_2$ partitions and consumed-charge trends, propagate their graphical uncertainty, and represent each product by a non-negative spectral channel probability. Because the source does not provide acquisition-time-resolved partial currents or state-to-state transition matrices, the absolute transfer scale is retained as a declared sensitivity parameter. The resulting operator is data-informed but not a measured absolute quantum-yield spectrum. Supplementary Note~7 formalizes this evidence-to-operator mapping, and Supplementary Table~7 records the raw-data boundary, declared transfer scale, finite design domain, and benchmark values that define the case-study claim boundary.

\begin{figure}[t]
\centering
\includegraphics[width=\textwidth]{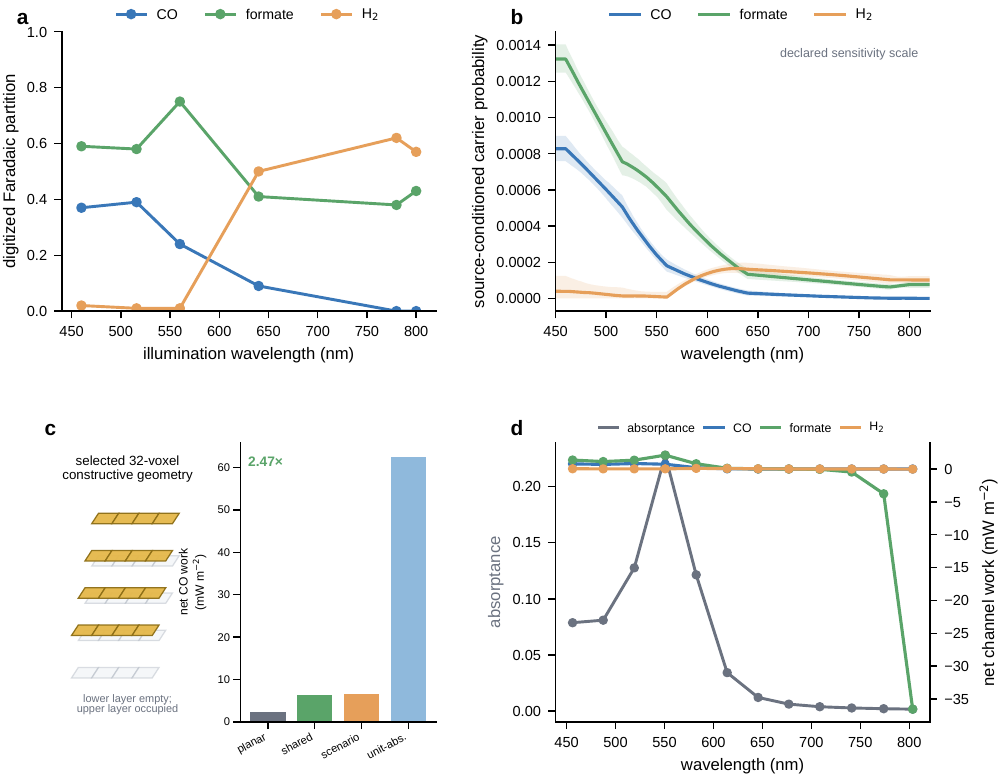}
\caption{\textbf{Source-conditioned gold on p-type gallium nitride illustration.} \textbf{a}, Wavelength-resolved Faradaic product partition digitized from public operando measurements performed at constant absorbed optical power. Blue, green, and orange traces denote carbon monoxide (CO), formate, and molecular hydrogen (H$_2$), respectively. The panel constrains product selectivity but does not by itself identify an absolute partial-current responsivity. \textbf{b}, Non-negative source-conditioned carrier-probability envelopes obtained by combining the product partitions in panel \textbf{a} with the relative consumed-charge-per-absorbed-power trend. Shaded intervals propagate graphical source uncertainty, and the absolute transfer scale remains a declared sensitivity parameter rather than an inferred quantum yield. \textbf{c}, Constructive one-common-structure calculation for a 32-voxel Au/p-GaN design domain. The selected upper and lower voxel layers are shown beside bars comparing the same-volume planar response, the shared three-dimensional geometry, a scenario-specific heuristic and a deliberately loose unit-absorptance channel bound. Within the declared model, the shared geometry gives 2.47 times the planar carbon-monoxide work and remains below both diagnostics. \textbf{d}, Independent twelve-bin spectral reevaluation of the selected shared geometry. The grey left-axis curve is absorptance, whereas the blue, green, and orange right-axis curves are net CO, formate, and H$_2$ channel work. Their differing peaks show that optical absorption and useful chemistry are not interchangeable objectives; negative long-wavelength formate work results from retaining the reciprocal reverse term.}
\label{fig:case}
\end{figure}

Figure~\ref{fig:case}a shows the digitized Faradaic partition. Formate dominates near 560~nm wavelength, CO decreases across the measured wavelength range, and H$_2$ rises sharply above approximately 600~nm wavelength. Figure~\ref{fig:case}b multiplies these partitions by the relative consumed-charge envelope and propagates the graphical uncertainty to obtain positive source-conditioned carrier-probability envelopes; Supplementary Figure~6d resolves the wavelength-dependent width of this propagated interval, while the absolute vertical scale remains declared rather than inferred. Figure~\ref{fig:case}c applies the operator to a $4\times4\times2$ binary design domain containing 32 Au voxels over a $70\times70\times14$~nm$^3$ region and a 160-nm normalization pitch. A multistart search over 12,779 candidate masks finds a shared structure with refined net CO work of \SI{6.576e-3}{W.m^{-2}}. The same-volume planar comparator gives \SI{2.614e-3}{W.m^{-2}}, so the constructive three-dimensional geometry is 2.47 times better within the stated model. Allowing scenario-specific masks gives \SI{6.904e-3}{W.m^{-2}}, a 6.65\% one-structure penalty, whereas a deliberately loose unit-absorptance bound remains much larger. The six-to-twelve-bin reevaluation and planar comparison are shown in Supplementary Figure~4b.

Figure~\ref{fig:case}d independently reevaluates the selected mask on 12 spectral bins. The grey curve is absorptance, and the product-coloured curves are net channel work. Their mismatch demonstrates that an optical peak need does not coincide with a productive spectral channel. The formate channel becomes negative at the reddest point, because the reciprocal reverse term exceeds forward work as the chemical load approaches the photon energy. These quantities are constructive lower and upper benchmarks for a finite design space; neither the selected voxel mask nor the absolute channel scale is asserted to be a globally optimal or directly fabricable catalyst. Sensitivity to the reflected Green tensor and normalization pitch is analyzed within Supplementary Note~7 rather than being folded into the headline geometry comparison.

\subsection*{Bias- and separation-complete energy ledger}
A biased photoelectrode may produce more product under illumination, while consuming more external electrical work than the additional chemical free energy it stores. The conditional operating point combines the locally detailed-balanced six-state kinetic family specified in Supplementary Note~8 and Supplementary Table~8 with the one-dimensional heat- and mass-transfer closure derived in Supplementary Note~9. Then, we evaluate an incremental light-minus-dark ledger,
\begin{equation}
\Delta P_{\rm net}=\sum_r\Delta\mathcal J_r\Delta_rG-\Delta P_{\rm sep}-\Delta P_{\rm bias}.
\label{eq:ledger}
\end{equation}
The symbol $\Delta$ denotes the illuminated-minus-dark increment. Thus, $\Delta\mathcal J_r$ is the incremental molar flux of product $r$, and every term in Eq.~\ref{eq:ledger} is an areal power density. For a gas product with outlet mole fraction $y_r$, the reversible separation contribution is $RT\mathcal J_r\ln(1/y_r)$; for a dissolved product concentrated from feed concentration $c_{\rm feed}$ to target concentration $c_{\rm target}$, it is $RT\mathcal J_r\ln(c_{\rm target}/c_{\rm feed})$, where $R$ is the molar gas constant and $T$ is the separation temperature. The electrical term is $|\Delta j|V_{\rm cell}$, where $\Delta j$ is incremental current density and $V_{\rm cell}$ is full-cell voltage, under the conservative measured-total-current convention.

\begin{figure}[t]
\centering
\includegraphics[width=\textwidth]{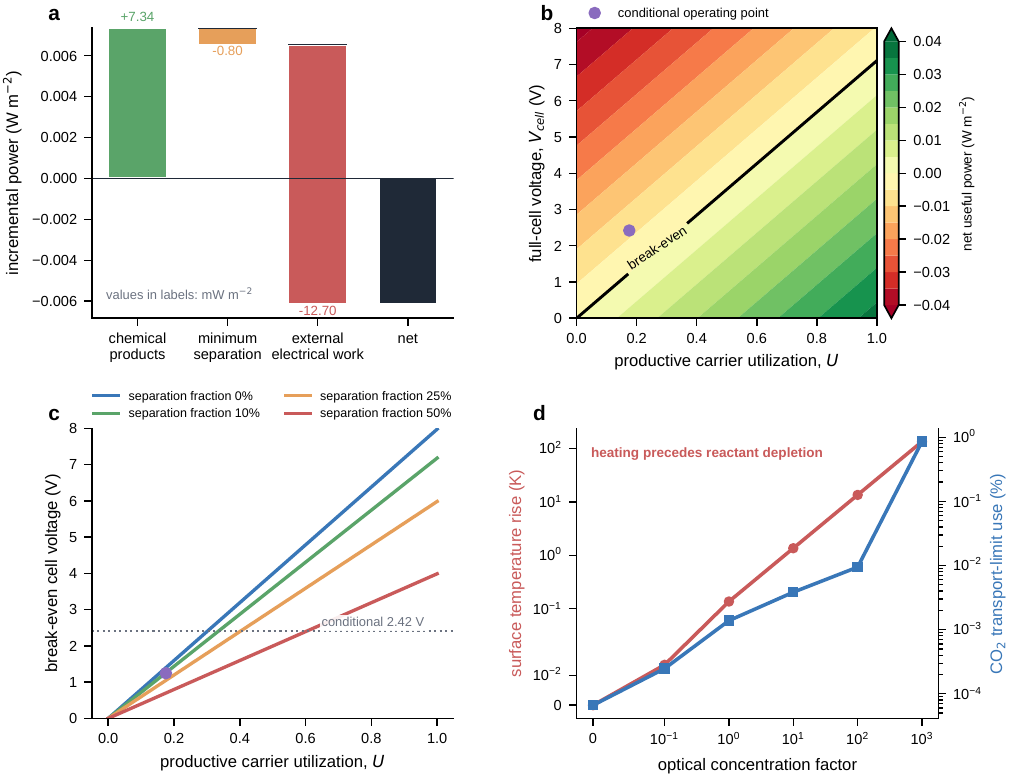}
\caption{\textbf{Bias-, separation-, heat-, and mass-transfer-complete system accounting.} \textbf{a}, Conditional incremental light-minus-dark power ledger. The green bar is additional chemical-product Gibbs free energy, the orange bar is ideal minimum product-separation work, the red bar is externally supplied electrical work, and the dark-grey bar is their net sum according to Eq.~\ref{eq:ledger}. Labels report milliwatts per square metre; the negative net illustrates that illumination-enhanced product formation is not necessarily net solar-energy conversion. \textbf{b}, Break-even phase map versus productive carrier utilization $U$ and full-cell voltage $V_{\rm cell}$. Colour gives net useful power, the black contour satisfies $P_{\rm net}=0$, the labelled region below the contour is energetically favourable, and the purple marker identifies the conditional case-study point. \textbf{c}, Break-even voltage $V_{\rm BE}$ versus $U$ for separation-work fractions of 0, 10, 25 and 50\%. The curves quantify how productive carrier use expands the allowable electrical-work budget and how separation burden contracts it. The purple marker is the fixed-kinetics conditional value; the horizontal dotted line is the conditional 2.42-V operating assumption. \textbf{d}, Continuum heat- and mass-transfer sensitivity versus optical concentration factor. Red circles give catalyst-surface temperature rise on the left logarithmic axis, and blue squares give the fraction of the carbon-dioxide diffusion limit on the right logarithmic axis. Under the stated one-dimensional closure, heating becomes a leading constraint before reactant depletion.}
\label{fig:ledger}
\end{figure}

Figure~\ref{fig:ledger}a applies this ledger to an explicitly conditional Au/p-GaN operating point. A locally detailed-balanced six-state reaction network gives a productive photoelectron utilization of 17.54\%; Supplementary Figure~5c resolves the associated product fractions, and Supplementary Figure~5d shows how the illustrative one-sun product partition closes against the delivered photoelectron flux. The incremental chemical output is \SI{7.343e-3}{W.m^{-2}}, the ideal minimum separation work is \SI{0.802e-3}{W.m^{-2}}, and the external electrical work at 2.42~V is \SI{12.701e-3}{W.m^{-2}}, leaving a conservative net of $-\SI{6.160e-3}{W.m^{-2}}$. Figure~\ref{fig:ledger}b generalizes the result as a phase map in productive carrier utilization $U$ and full-cell voltage $V_{\rm cell}$; the black contour separates positive and negative net useful power, and the conditional point lies above the break-even line. Figure~\ref{fig:ledger}c resolves the break-even voltage for separation burdens from 0 to 50\%. For a generic delivered chemical-power limit $P_{\rm ch}^{0}$, defined as the gross product free-energy rate at unit productive utilization before separation and bias, let $U$ be the productive carrier-utilization fraction, $f_{\rm sep}$ the fraction of productive chemical power consumed by minimum separation, and $j_{\rm ext}$ the incremental externally supplied current density. Then
\begin{equation}
P_{\rm net}=UP_{\rm ch}^{0}(1-f_{\rm sep})-j_{\rm ext}V_{\rm cell},
\qquad
V_{\rm BE}=\frac{UP_{\rm ch}^{0}(1-f_{\rm sep})}{j_{\rm ext}}.
\label{eq:breakEven}
\end{equation}
Here, $P_{\rm net}$ is net useful power density and $V_{\rm BE}$ is the full-cell voltage at which it vanishes. The fixed-kinetics conditional break-even voltage is 1.246~V. Supplementary Note~10 derives the gas- and dissolved-product separation terms used in this construction; Supplementary Figure~7a resolves their dependence on dilution and target concentration, and Supplementary Table~9 gives the complete numerical ledger.

Figure~\ref{fig:ledger}d couples the residual heat and product fluxes to a one-dimensional continuum model. At one sun, the conditional surface-temperature rise is 0.136~K and the carbon flux uses only $1.38\times10^{-5}$ of the estimated CO$_2$ transport limit. The corresponding temperature, carbon-dioxide depletion, and product-concentration profiles are shown in Supplementary Figure~6a--c. At 1000-fold concentration, the calculated temperature rise reaches 136~K, while CO$_2$ transport-limit use remains below 1\%. Thus, heating precedes reactant depletion under the stated closure, although the precise high-concentration temperature is outside the range, where constant material properties and negligible convection remain reliable. Supplementary Figure~7b shows how the same conditional network redistributes product selectivity with optical concentration.

\section*{Discussion}
The theory closes a gap between three established traditions. Photochemical detailed balance constrains the free energy that radiation can store\cite{Ross1967,RossHsiao1977,Fingerhut2010}; PEC limits couple an absorber to catalytic and transport losses\cite{Fountaine2016}; and nanophotonic bounds constrain the response attainable by passive matter\cite{Miller2014,ZhangMonticoneMiller2023,Shim2024}. Plasmonic solar chemistry requires all three, because its absorber, carrier generator, heat source, and catalytic interface can be the same nanostructure. Treating these roles independently can violate energy conservation or assign mutually incompatible spectral responses to one catalyst.

Several results are broadly transferable. The chemical diode law supplies an SQ-like analytic benchmark for any tightly coupled solar chemical cycle and recovers both a variable-load photovoltaic limit and a fixed-load PEC limit. The one-structure calculations establish that per-frequency optima are not a legitimate broadband limit: penalties of 6.6--33.8\% arise in finite but qualitatively different benchmarks, and cross-frequency correlations reduce certified upper bounds by 21--56\%. These percentages are not universal constants, but they show that realizability can be a leading-order loss. The exact $g\tau$ law likewise identifies a material-independent multi-electron bottleneck. At one sun, storing and spatially concentrating redox equivalents can be more valuable than modestly increasing extinction.

The formulation also clarifies what can and cannot be claimed from incomplete microscopic data. A positive-semidefinite source-conditioned operator can incorporate wavelength-resolved selectivity and uncertainty without pretending that digitized charge is an absolute partial-current responsivity. An admissible locally detailed-balanced kinetic family can constrain product fluxes without presenting fitted barrier offsets as first-principles transition states. The Au/p-GaN device calculations are therefore an illustration of how evidence narrows a universal feasible set. They are not an asserted efficiency of Au/p-GaN device, and the finite voxel designs are not universal material bounds.

More generally, a useful plasmonic catalyst should not be judged by absorption, hot-carrier yield, rate enhancement, or temperature rise in isolation. The relevant benchmark is the net Gibbs free energy retained after reciprocal back reactions, incomplete carrier use, heat and mass transport, applied electrical work, and product separation, under an electromagnetic response that one passive nanostructure can actually realize. The framework developed here provides that benchmark and a common language for comparing direct, hot-carrier, and photothermal routes to solar chemistry.

\section*{Methods}
\subsection*{Solar and ambient photon spectra}
The incident spectrum was the ASTM International G173 global-tilt reference spectrum (ASTM G173) between 280 and 4000~nm wavelength. Spectral irradiance $I_\lambda$ was converted to photon flux as $\Phi_\odot(\lambda)=I_\lambda\lambda/(hc)$. The ambient reverse photon flux was the 300-K hemispherical blackbody flux,
\begin{equation}
\Phi_0(\lambda,T_0)=\frac{2\pi c}{\lambda^4}\left[\exp\!\left(\frac{hc}{\lambda\kB T_0}\right)-1\right]^{-1},
\label{eq:blackbody}
\end{equation}
where $\Phi_0$ is the hemispherical ambient photon flux density per unit wavelength, $\lambda$ is vacuum wavelength, $h$ is Planck's constant, $c$ is the vacuum speed of light, and $T_0$ is ambient temperature. The solar photon flux density is $\Phi_\odot(\lambda)=I_\lambda\lambda/(hc)$, where $I_\lambda$ is spectral irradiance. Numerical integration used the native ASTM G173 wavelength grid.

\subsection*{Reduced chemical diode}
Let $E=hc/\lambda$ and convert per-wavelength photon fluxes to per-energy fluxes with $\Phi(E)=\Phi(\lambda)|\dd\lambda/\dd E|$. For a dimensionless action spectrum $a(E)$, $G=\int_0^\infty a(E)[\Phi_\odot(E)-\Phi_0(E,T_0)]\dd E$ and $R_0=\int_0^\infty a(E)\Phi_0(E,T_0)\dd E$. Equations~\ref{eq:chemicaldiode}--\ref{eq:lambert} were evaluated analytically. The step action was $a(E)=\Theta(E-E_{\rm th})$, and the Fowler action was Eq.~\ref{eq:fowler}. The fixed-load photoelectrochemical reduction evaluated $P=\mu\max[J(\mu),0]$ at $\mu=1.23$~eV, with electronvolt values converted to joules before multiplication by flux.

\subsection*{Matrix-oscillator and one-structure bounds}
The finite matrix-oscillator program used PSD angular oscillator atoms and imposed the two integrated constraints in Eq.~\ref{eq:sumrules}, together with a per-scenario absorptance cap. Direct, hot-carrier and heat objectives were linear functionals of the oscillator weights; a reversible one-node thermal-exergy calculation converted residual heat into an outer work bound. One-common-eigenbasis and one-axis restrictions were imposed by sharing the oscillator orientation across frequencies. The operator definitions are derived in Supplementary Note~4, and the numerical oscillator budgets are reported in Supplementary Table~3.

For exactly enumerable benchmarks, binary material masks were solved with a volume-integral equation. Scenario-specific optima were obtained by choosing a different mask for each wavelength and polarization. Shared optima used one mask for all scenarios. Lifted SDP relaxations were evaluated with and without off-diagonal cross-scenario constraints. A dual largest-eigenvalue correction ensured positive-semidefinite slack, so reported dual objectives are certified upper bounds for the finite problems. Supplementary Note~5 specifies the lifted constraints and dual correction, Supplementary Figure~3 records the numerical certificates, and Supplementary Figure~4 tests spectral refinement independently.

\subsection*{Quasistatic material screen}
Ag, Au and Rh nanoparticle responses were computed from measured or tabulated complex permittivities using the exact finite-loss quasistatic ellipsoid response and the material extinction sum-rule limit\cite{Miller2014}. Sphere, optimized fixed-spheroid, wavelength-wise spheroid, and material-bound hierarchies were compared at a 1-eV Fowler threshold and 5-nm equivalent metal thickness. Extinction was treated as potentially useful absorption, making this an intentionally optimistic screen. Supplementary Figure~2 resolves the finite-loss correction and geometry hierarchy, while Supplementary Table~4 provides the values used to construct the screen.

\subsection*{Multi-electron network}
The sequential-reset process contained states 0 through $n-1$. Productive carrier arrival advanced $i\to i+1$ at rate $g$, with $n-1\to0$ completing a product. Every occupied intermediate reset to zero at rate $1/\tau$. The stationary distribution and Eq.~\ref{eq:multielectron} follow from the master-equation solution in Supplementary Note~6. Supplementary Figure~5a,b supplies the independent generator check, and Supplementary Table~6 gives the utilization roots reported in Figure~4b.

\subsection*{Source-conditioned operator and finite Au/p-GaN device model}
Wavelength-dependent product partitions and consumed-charge trends were digitized from the public Au/p-GaN device operando study\cite{Kiani2026}. Uncertainty bands propagated the reported graphical uncertainty through monotone interpolation. Non-negative product-channel probabilities were normalized by a declared peak-transfer sensitivity; no state-resolved density-functional-theory or many-body $GW$ transition rows and no absolute time-resolved partial-current responsivities were inferred. Supplementary Table~7 records this source boundary. The optical model used Johnson--Christy Au data, a measured p-GaN optical response, an aqueous upper medium, retarded dipole interactions, and a planar reflected Green tensor, as specified in Supplementary Note~7. The 32-voxel design was searched by multistart constructive optimization and independently refined on a 12-bin grid; Supplementary Figure~4b gives the refinement check, and Supplementary Figure~6d gives the propagated operator uncertainty.

\subsection*{Conditional kinetics, heat/mass transfer and system ledger}
The conditional case used the six surface states and reversible CO, formate, and H$_2$ pathways defined in Supplementary Note~8 and Supplementary Table~8; all forward/backward rate pairs obeyed Eq.~\ref{eq:ldb}, and the fitted barrier offsets were retained only as source-conditioned sensitivity parameters. One-dimensional diffusion in a stagnant electrolyte boundary layer was coupled to heat conduction and an external heat-transfer coefficient according to Supplementary Note~9, with the resulting profiles shown in Supplementary Figure~6a--c. The incremental ledger used Eq.~\ref{eq:ledger}; Supplementary Note~10 and Supplementary Table~9 define the ideal gas/dissolved separation terms and the reference electrical-work convention, while Supplementary Figure~7 displays their dilution and concentration sensitivities.

\subsection*{Numerical implementation and verification}
All main and Supplementary figures were generated from numerical tables derived from the cited source data and the calculations described above. Independent checks include the ASTM International G173 irradiance integral, Lambert-$W$ stationarity, containment of spheroidal responses by the finite-loss material limit, exact-enumeration/semidefinite-programming ordering, Markov-generator agreement, spectral refinement, local detailed balance, network stationarity, and arithmetic closure of the energy ledger. Supplementary Note~11 documents the implementation and validation logic, and Supplementary Table~10 reports the corresponding numerical residuals and checkpoint values.

\subsection*{Experimental validation protocol}
Supplementary Note~12 specifies a prospective validation and parameter-identification protocol for calibrating the channel operators with wavelength-resolved operando measurements. The common electrode/cell geometry is shown in Supplementary Figure~8a, with the sample and device-stack record defined in Supplementary Table~11. The equal-absorbed-power wavelength calibration in Supplementary Figure~8b is paired with the optical and thermal record in Supplementary Table~12. Product-resolved photo-scanning electrochemical microscopy and the complementary mechanism-discrimination tests in Supplementary Figure~8c are specified in Supplementary Table~13. Finally, Supplementary Figure~8d traces native data through calibration, absorbed-power normalization, and channel-operator construction, while Supplementary Table~14 sets the quality-control, uncertainty and mass-balance criteria for that workflow.

\subsection*{Data availability}
The numerical data supporting the findings of this study are contained in the Article and Supplementary Information and are available from the corresponding author upon reasonable request. Digitized public Au/p-GaN device observables are distinguished from the derived channel-operator data.

\subsection*{Code availability}
The code used to generate the numerical results and figures is available from the corresponding author upon reasonable request.

\section*{Acknowledgements}
This work is supported by National Research Foundation of Korea (NRF-RS-2023-00272363 and RS-2026-25621181) and Korea University grant.

\section*{Author contributions}
S.L. conceived the study, developed the theoretical framework, performed the calculations, and wrote the manuscript.

\section*{Competing interests}
The author declares no competing interests.

\end{document}


\maketitle
\tableofcontents
\clearpage

\noteheading{1. Scope, notation and hierarchy of admissible converters}

The purpose of this Supplementary Information is to make explicit the assumptions, reductions, numerical certificates and source boundaries underlying the main text. We refer to the resulting variational limit as the plasmonic solar-chemical detailed-balance (PSCD) limit. The theory separates three levels of statement. First, the \emph{universal} level follows from energy conservation, microscopic reversibility, non-negative entropy production, and electromagnetic passivity. Second, the \emph{mechanism-constrained} level introduces direct interfacial excitation, nonequilibrium hot-carrier generation, carrier survival, and photothermal conversion. Third, the \emph{source-conditioned} level narrows the admissible operators using wavelength-resolved public observations. The third level is an illustration of how data contract a universal feasible set; it is not presented as a first-principles prediction of gold/p-type gallium nitride (Au/p-GaN). The source-conditioned example tracks carbon dioxide (CO$_2$), carbon monoxide (CO), formate, and molecular hydrogen (H$_2$); these chemical abbreviations are used consistently below.

\textbf{Symbol and unit conventions.} Unless an equation states otherwise, optical and system quantities denoted by $P$ are areal power densities in W~m$^{-2}$. Probability currents $J_{ij}^{(a)}$ and the per-reservoir completion rate $J_n$ have units of s$^{-1}$; $J(\mu)$ is a reduced cycle-completion flux in the normalization used for the chemical diode; $\mathcal J_r$ and $\mathcal J_p$ are areal molar reaction/product fluxes in mol~m$^{-2}$~s$^{-1}$; $j$ denotes electrical current density in A~m$^{-2}$; and $\mathcal N_i$ denotes diffusive molar flux. Energies inside Boltzmann factors are per event in joules; values quoted in electronvolts are converted with the elementary charge $q$. Molar free energies are used only when multiplied by molar fluxes. The superscript $\star$ denotes a supremum, the superscript $(0)$ denotes the initial post-absorption branch before relaxation, $\dagger$ denotes Hermitian conjugation, and $\preceq$/$\succeq$ denote the Loewner order for Hermitian matrices. Indices $i,j$ label internal states, $a$ labels a mediating reservoir, $\ell$ labels chemical species, $r$ labels useful reactions, $p$ labels products, $c$ labels optical dissipation channels, and $s$ labels optical scenarios.

For an incident mode $m=(\omega,\Omega,\sigma)$, where $\omega$ is angular frequency, $\Omega$ is propagation direction, and $\sigma$ is polarization, $P_{\rm ext}$, $P_{\rm sca}$ and $P_{\rm abs}$ denote extinction, scattering and absorption power, respectively, and obey
\begin{equation}
P_{\rm ext}(m)=P_{\rm sca}(m)+P_{\rm abs}(m).
\label{eq:Sextinction}
\end{equation}
The first irreversible partition of absorbed power is
\begin{equation}
P_{\rm abs}(m)=P_{\rm dir}^{(0)}(m)+P_{\rm hc}^{(0)}(m)+P_{\rm ph}^{(0)}(m),
\label{eq:Spartition}
\end{equation}
where the three terms denote direct metal--adsorbate excitation, nonequilibrium intrametal carrier generation, and initially thermalized power. The superscript $(0)$ identifies the initial branching before subsequent relaxation or reaction. Unsuccessful direct and carrier-mediated events subsequently enter
\begin{equation}
P_{\rm heat}=P_{\rm ph}^{(0)}+Q_{\rm dir,loss}+Q_{\rm hc,loss}.
\label{eq:Sheat}
\end{equation}
Here, $Q_{\rm dir,loss}$ and $Q_{\rm hc,loss}$ are the direct- and hot-carrier-channel powers that fail to perform chemical work and subsequently thermalize. Equations~\eqref{eq:Spartition} and \eqref{eq:Sheat} are bookkeeping constraints, not assumptions about which mechanism dominates. They prevent the same absorbed energy from being counted independently as both quantum chemical work and heat.

The nested admissible sets used throughout the manuscript are
\begin{equation}
\mathcal C_{\rm data}\subseteq\mathcal C_{\rm mech}\subseteq\mathcal C_{\rm univ},
\qquad
\eta_{\rm data}^{\star}\leq\eta_{\rm mech}^{\star}\leq\eta_{\rm electromagnetic}^{\star}\leq\eta_{\rm ideal\,photochemical}^{\star}.
\label{eq:Snested}
\end{equation}
Here, $\mathcal C_{\rm univ}$ is the universal positive-semidefinite channel set, $\mathcal C_{\rm mech}$ additionally imposes carrier-generation, survival and transfer restrictions, and $\mathcal C_{\rm data}$ additionally imposes source observations. The four efficiencies are the corresponding source-conditioned, mechanism-constrained, electromagnetic-only, and ideal-photochemical suprema. A transition matrix obtained from electronic-structure calculations would further contract $\mathcal C_{\rm data}$, but is not required to define the preceding suprema. This distinction defines the theory-first scope of the work.

\begin{table}[H]
\centering
\caption{Core notation used in the detailed-balance framework.}
\label{tab:notation}
\begin{tabularx}{\textwidth}{>{\raggedright\arraybackslash}p{0.19\textwidth}X}
\toprule
Symbol & Definition \\
\midrule
$m=(\omega,\Omega,\sigma)$ & Radiation mode specified by angular frequency, propagation direction, and polarization. \\
$P_{\rm dir}^{(0)},P_{\rm hc}^{(0)},P_{\rm ph}^{(0)}$ & Mutually exclusive first branches of absorbed optical power: direct interfacial excitation, intrametal nonequilibrium-carrier generation, and initially thermalized power. \\
$J_{ij}^{(a)}$ & Probability current for transition $i\rightarrow j$ mediated by reservoir or channel $a$. \\
$\Delta_rG$ & Gibbs free energy stored per completed useful reaction at the reference state. \\
$G,R_0$ & Excess useful photogeneration rate and equilibrium reverse-cycle prefactor in the reduced chemical diode. \\
$\mu_{\rm oc},\mu_{\rm mp}$ & Stall and maximum-power chemical free energies per completed cycle. \\
$g,\tau,n$ & Useful carrier arrival rate at one accumulation reservoir, intermediate lifetime, and required carrier count. \\
$\mathsf T(\omega),\mathsf X(\omega)$ & Scattering response matrix and positive-semidefinite matrix-oscillator spectral measure. \\
$P_{\rm bias},P_{\rm sep}$ & Externally supplied electrical power and minimum reversible product-separation power. \\
\bottomrule
\end{tabularx}
\end{table}

\noteheading{2. Local detailed balance and the universal efficiency theorem}

Let $p_i$ be the dimensionless probability of a catalyst--adsorbate--charge-storage state $i$. An elementary transition $i\rightarrow j$ mediated by reservoir or channel $a$ has probability current
\begin{equation}
J_{ij}^{(a)}=k_{ij}^{(a)}p_i-k_{ji}^{(a)}p_j,
\qquad J_{ji}^{(a)}=-J_{ij}^{(a)}.
\label{eq:Sedgecurrent}
\end{equation}
Here, $k_{ij}^{(a)}$ and $k_{ji}^{(a)}$ are forward and reverse rate constants, and positive $J_{ij}^{(a)}$ denotes net flow from $i$ to $j$. At steady state,
\begin{equation}
\sum_{j,a}J_{ji}^{(a)}=0
\label{eq:Ssteady}
\end{equation}
for every state. Thermal and electrochemical steps satisfy local detailed balance,
\begin{equation}
\ln\frac{k_{ij}^{(a)}}{k_{ji}^{(a)}}
=-\frac{\Delta E_{ij}-\sum_{\ell}\mu_{\ell}\Delta N_{\ell,ij}}{\kB T_a}.
\label{eq:Sldb}
\end{equation}
In Eq.~\eqref{eq:Sldb}, $\Delta E_{ij}=E_j-E_i$ is the system internal-energy change, $\Delta N_{\ell,ij}=N_{\ell,j}-N_{\ell,i}$ is the particle-number change of species $\ell$, $\mu_{\ell}$ is its chemical potential, $T_a$ is the reservoir temperature, and $k_{\mathrm B}$ is Boltzmann's constant. For a nonthermal electronic reservoir, the two rates are instead evaluated from the actual energy-resolved occupation, Pauli factors, and transition kernel; the framework does not impose an artificial single electron temperature or quasi-Fermi splitting on a nonthermal metal.

For each bidirectional edge, define $x=k_{ij}^{(a)}p_i$ and $y=k_{ji}^{(a)}p_j$. Its entropy-production contribution is
\begin{equation}
\dot S_{ij}^{(a)}=\kB(x-y)\ln\frac{x}{y}\geq0,
\label{eq:Sedgeentropy}
\end{equation}
because $(x-y)\ln(x/y)\geq0$ for all positive $x$ and $y$. The quantities $x$ and $y$ are the forward and reverse one-way probability fluxes on the same edge. Summing independent edges, with $i<j$ counting each bidirectional edge once, gives
\begin{equation}
\dot S_{\rm gen}=\kB\sum_{i<j,a}J_{ij}^{(a)}
\ln\frac{k_{ij}^{(a)}p_i}{k_{ji}^{(a)}p_j}\geq0.
\label{eq:Sentropy}
\end{equation}
This proof remains valid for arbitrarily branched reaction networks as long as each reservoir-mediated edge is represented by its forward and reverse processes.

Let $d_{ij}^{(r)}$ be the signed stoichiometric incidence coefficient that counts completed events of useful reaction $r$ when edge $i\leftrightarrow j$ carries unit net current. The net product flux is
\begin{equation}
\mathcal J_r=\sum_{i<j,a}d_{ij}^{(r)}J_{ij}^{(a)},
\label{eq:Sproductflux}
\end{equation}
and useful chemical power density is $P_{\rm ch}=\sum_r\mathcal J_r\Delta_rG$, where $\mathcal J_r$ is the net areal molar reaction flux and $\Delta_rG>0$ is molar Gibbs free energy stored relative to the declared reference state. Activation barriers are not counted as stored output. The universal plasmonic solar-chemical limit is therefore
\begin{equation}
\eta_{\rm PSCD}^{\star}=
\frac{1}{P_{\odot,\rm inc}}
\sup_{\mathcal F}
\left[
\sum_r\mathcal J_r\Delta_rG-P_{\rm bias}-P_{\rm sep}
\right],
\label{eq:Smaster}
\end{equation}
where $P_{\odot,\rm inc}$ is incident solar power density, $P_{\rm bias}$ and $P_{\rm sep}$ are external electrical and minimum reversible separation power densities, and $\mathcal F$ enforces Maxwell's equations; passivity, causality, and sum rules; one common geometry across all optical scenarios; Eqs.~\eqref{eq:Spartition}, \eqref{eq:Ssteady}, \eqref{eq:Sldb} and \eqref{eq:Sentropy}; atomic and site conservation; multi-electron accumulation; and the thermal and mass-transfer constraints stated below. Equation~\eqref{eq:Smaster} is a variational definition. The finite numerical examples in the manuscript are inner constructions or outer relaxations of subsets of $\mathcal F$ and are labelled accordingly.

The independent-rate organization is important. An apparently additive loss decomposition can repeat the same absorptance, escape probability, or relaxation pathway in multiple terms. The safer form is a network of independent forward, reverse, and parasitic rates whose currents enter Eqs.~\eqref{eq:Sedgecurrent}--\eqref{eq:Sentropy}. This follows the same logic used to remove duplicate optical-loss terms in photovoltaic light-management thermodynamics\cite{Rau2014}.

\noteheading{3. Chemical diode theorem and reductions to established limits}

\subsection*{3.1 Reduced tightly coupled cycle}

Consider a tightly coupled useful cycle in which each completed forward cycle stores free energy $\mu$. Let $G$ be the useful excess photogeneration rate and $R_0$ the equilibrium reverse-cycle prefactor. Raising the product chemical potential increases the reciprocal reverse current by $\exp(\mu/\kB T_0)$. The net cycle rate is
\begin{equation}
J(\mu)=G-R_0\left[\exp\left(\frac{\mu}{\kB T_0}\right)-1\right].
\label{eq:Sdiode}
\end{equation}
This is the chemical analogue of the ideal diode law. At stall, $J=0$, giving
\begin{equation}
\mu_{\rm oc}=\kB T_0\ln\left(1+\frac{G}{R_0}\right).
\label{eq:Smuoc}
\end{equation}
For a fixed target reaction $\Delta_rG$, the useful power is
\begin{equation}
P_r=\Delta_rG\left\{G-R_0\left[\exp\left(\frac{\Delta_rG}{\kB T_0}\right)-1\right]\right\}_{+}.
\label{eq:Sfixedpower}
\end{equation}
where $\{z\}_{+}=\max(z,0)$ prevents a negative forward product flux from being counted as useful output.

\subsection*{3.2 Exact maximum-power load}

Set $\xi=\mu/(\kB T_0)$ and $\gamma=G/R_0$. Maximizing
\begin{equation}
P(\xi)=\kB T_0R_0\xi\left[\gamma+1-e^\xi\right],
\label{eq:SPx}
\end{equation}
requires
\begin{equation}
1+\gamma=e^\xi(1+\xi).
\label{eq:Sstationary}
\end{equation}
With $u=1+\xi$, Eq.~\eqref{eq:Sstationary} becomes $ue^u=e(1+\gamma)$. Hence
\begin{equation}
\boxed{\mu_{\rm mp}=\kB T_0\left[W\!\left(e(1+\gamma)\right)-1\right]},
\label{eq:Slambert}
\end{equation}
where $W$ is the principal branch of the Lambert $W$ function, $e$ is Euler's number, $\xi=\mu/(\kB T_0)$ is dimensionless, and $\gamma=G/R_0$ is the generation-to-reverse-rate ratio. At this point,
\begin{equation}
J_{\rm mp}=R_0\xi e^\xi,
\qquad
P_{\rm mp}=\kB T_0R_0\xi^2e^\xi,
\label{eq:Smp}
\end{equation}
where $J_{\rm mp}$ and $P_{\rm mp}$ are the cycle flux and chemical power at the maximum-power load. The chemical fill factor, defined as $FF_{\rm ch}=P_{\rm mp}/(G\mu_{\rm oc})$, is
\begin{equation}
FF_{\rm ch}=\frac{\xi^2e^\xi}{\gamma\ln(1+\gamma)}.
\label{eq:Sfillfactor}
\end{equation}

For a dimensionless action spectrum $a(E)$, where $E$ is photon energy, $\Phi_{\odot}(E)$ and $\Phi_0(E,T_0)$ are solar and ambient photon flux densities per unit energy, and $\dd E$ is the energy differential,
\begin{equation}
G=\int_0^\infty a(E)\left[\Phi_{\odot}(E)-\Phi_0(E,T_0)\right]\dd E,
\qquad
R_0=\int_0^\infty a(E)\Phi_0(E,T_0)\dd E.
\label{eq:SGR}
\end{equation}
The two actions used for numerical reductions are
\begin{equation}
a_{\rm step}(E)=\Theta(E-E_{\rm th}),
\qquad
a_{\rm F}(E)=\left(\frac{E-E_{\rm th}}{E}\right)^2\Theta(E-E_{\rm th}).
\label{eq:Sactions}
\end{equation}
Here, $E_{\rm th}$ is the threshold energy and $\Theta$ is the Heaviside step function. The first is a unit-yield threshold absorber. The second is an aggregated Fowler escape law and should not be interpreted as a complete energy- and momentum-resolved metal-interface model.

\begin{table}[H]
\centering
\caption{Numerical reductions of the chemical-diode formulation. Efficiencies use the full ASTM International G173 (ASTM G173) global-tilt irradiance, 1000.3707 W m$^{-2}$.}
\label{tab:reductions}
\small
\begin{tabularx}{\textwidth}{>{\raggedright\arraybackslash}p{0.25\textwidth}>{\centering\arraybackslash}p{0.17\textwidth}>{\centering\arraybackslash}p{0.17\textwidth}X}
\toprule
Action/load model & Optimum threshold (eV) & Maximum efficiency (\%) & Interpretation \\
\midrule
Unit step, variable load & 1.34 & 33.68 & Single-threshold radiative envelope \\
Unit step, fixed $\mu=1.23$ eV & 1.59 & 30.58 & Ideal single-junction water-splitting reduction \\
Fowler action, variable load & 0.60 & 8.53 & Aggregated hot-carrier filtering limit \\
\bottomrule
\end{tabularx}
\end{table}

The variable-load step-action calculation reproduces the familiar one-sun single-threshold radiative envelope. Fixing $\mu=1.23$~eV gives a 30.58\% maximum at $E_{\rm th}=1.59$~eV, matching the ideal single-junction water-splitting limit reported in the unified photoelectrochemical (PEC) analysis of Fountaine \etal\cite{Fountaine2016}. That PEC analysis derives an inverse current--voltage equation, identifies operation when the generated voltage exactly supplies the reaction potential, and then introduces absorption, external radiative efficiency, resistance, and exchange-current losses. The present reduction recovers its ideal fixed-load limit, while the full theorem extends the converter space to plasmonic direct, hot-carrier, and thermal channels.

\begin{figure}[H]
\centering
\includegraphics[width=0.98\textwidth]{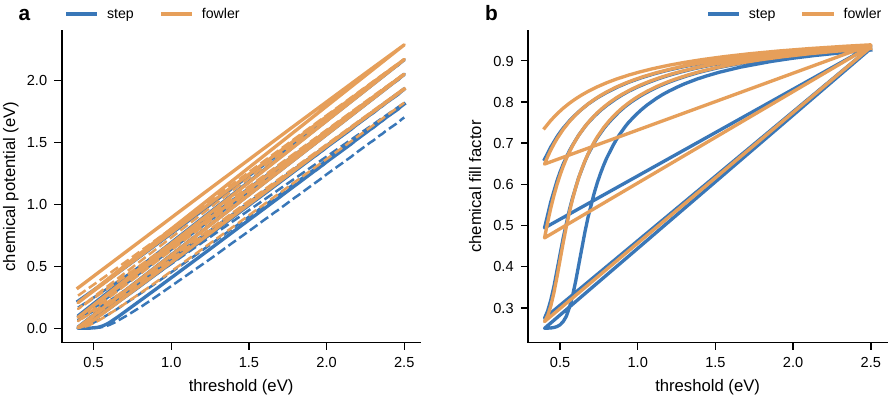}
\caption{\textbf{Chemical-potential and fill-factor diagnostics.} \textbf{a}, Chemical open-circuit, or stall, free energy $\mu_{\rm oc}$ (solid curves), and maximum-power free energy $\mu_{\rm mp}$ (dashed curves) for unit-step and Fowler action spectra over representative threshold energies. The separation between each solid and dashed curve is the load-matching penalty required to obtain finite power. \textbf{b}, Chemical fill factor $FF_{\rm ch}=P_{\rm mp}/(G\mu_{\rm oc})$ calculated from Eq.~\eqref{eq:Sfillfactor}. Each curve corresponds to a distinct threshold energy, and the approach toward unity at large $G/R_0$ reflects the increasingly square chemical-diode characteristic.}
\label{fig:Schemical}
\end{figure}

Supplementary Figure~\ref{fig:Schemical}a separates the stall and maximum-power loads for the two action spectra and shows that the maximum-power load remains below the reversible stall load for every threshold. Supplementary Figure~\ref{fig:Schemical}b converts the same solutions into a chemical fill factor and verifies the limiting behavior of Eq.~\eqref{eq:Sfillfactor}. Together with Supplementary Table~\ref{tab:reductions}, the two panels provide the numerical and analytic checks summarized in main-text Figure~2.

\subsection*{3.3 Limiting reductions}

The framework satisfies the following consistency reductions.
\begin{enumerate}[leftmargin=2em]
\item Setting a unit step action, treating $\mu=qV$, and retaining only reciprocal photon emission give the ideal single-junction photovoltaic diode of Shockley and Queisser\cite{Shockley1961}.
\item Fixing $\mu=1.23$~eV gives the ideal single-junction PEC water-splitting reduction discussed above\cite{Fountaine2016}.
\item Replacing the electromagnetic feasible set by an ideal molecular action spectrum produces a Ross-type photochemical-potential limit\cite{Ross1967,RossHsiao1977}.
\item Replacing the metal by an ideal thermalized semiconductor absorber with energy-selective extraction produces a semiconductor hot-carrier converter limit\cite{RossNozik1982,Takeda2022}.
\item Setting direct and hot-carrier chemical work to zero leaves a pure thermal-exergy optimization.
\end{enumerate}
These reductions do not imply that all physical assumptions of the corresponding models are identical; they establish that the new equations contain their idealized functional forms as restricted cases.

\noteheading{4. Passivity-constrained electromagnetic and channel-operator bounds}

\subsection*{4.1 Matrix-valued oscillator representation}

For optical scenario $s$, let $\mathbf e_s(\omega)$ be the incident-field coefficient vector and $\mathbf p_s(\omega)$ the induced polarization-current vector, related by
\begin{equation}
\mathbf p_s(\omega)=\mathsf T(\omega)\mathbf e_s(\omega).
\label{eq:STresponse}
\end{equation}
For a linear passive scatterer, the response can be represented by a positive-semidefinite (PSD) matrix-valued oscillator measure\cite{Zhang2023}. A convenient schematic form is
\begin{equation}
\mathsf T(\omega)=\lim_{\delta\rightarrow0^+}
\int_0^\infty
\frac{\mathsf X(\omega_i)+(\omega_i/\omega)\mathsf Y(\omega_i)}
{\omega_i^2-\omega^2-\mathrm{i}\delta\omega}\dd\omega_i,
\label{eq:Smatrixosc}
\end{equation}
where $\omega_i$ is oscillator frequency, $\delta\rightarrow0^+$ is the causal broadening limit, and $\mathrm{i}=\sqrt{-1}$ is the imaginary unit. The Hermitian matrix-valued spectral measures $\mathsf X$ and $\mathsf Y$ satisfy
\begin{equation}
\mathsf X(\omega_i)\succeq0,
\qquad
-\mathsf X(\omega_i)\preceq\mathsf Y(\omega_i)\preceq\mathsf X(\omega_i).
\label{eq:SXY}
\end{equation}
The high- and low-frequency sum rules used in the finite demonstration are
\begin{equation}
\int_0^\infty\mathsf X(\omega_i)\dd\omega_i\preceq\mathsf B_{\rm H},
\qquad
\int_0^\infty\frac{\mathsf X(\omega_i)}{\omega_i^2}\dd\omega_i\preceq\mathsf B_{\rm L}.
\label{eq:Ssumrules}
\end{equation}
The inequalities use the Loewner order. The matrices $\mathsf B_{\rm H}$ and $\mathsf B_{\rm L}$ encode the finite material volume, oscillator strength, and static response of the conditional model.

\subsection*{4.2 Positive channel operators}

Suppose the dissipative part of the electric susceptibility operator $\bm\chi$ is partitioned among optical-loss channels $c$,
\begin{equation}
\operatorname{Im}\bm\chi=\sum_c\operatorname{Im}\bm\chi_c,
\qquad
\operatorname{Im}\bm\chi_c\succeq0,
\label{eq:Ssuspartition}
\end{equation}
where $\operatorname{Im}$ denotes the Hermitian imaginary part. The corresponding channel work operator can be written
\begin{equation}
\mathsf W_c=\epsilon_0^{-1}\bm\chi^{-\dagger}
\left(\operatorname{Im}\bm\chi_c\right)\bm\chi^{-1}\succeq0,
\label{eq:SWc}
\end{equation}
where $\epsilon_0$ is vacuum permittivity, $\bm\chi^{-1}$ is the inverse susceptibility on the chosen finite response subspace, and $-\dagger$ denotes inverse Hermitian adjoint. The resulting $P_c^{(s)}(\omega)$ is the channel absorbed-power spectral density in W~m$^{-2}$ per unit angular frequency and is
\begin{equation}
P_c^{(s)}(\omega)=\frac{\omega}{2}\mathbf p_s^\dagger(\omega)\mathsf W_c(\omega)\mathbf p_s(\omega).
\label{eq:Schannel}
\end{equation}
With dimensionless yield kernel $Y_{c,r}(\omega,s)$, defined as useful reaction-$r$ events per absorbed quantum in channel $c$, useful generation is
\begin{equation}
G_r[\mathsf T]=\sum_s\int\frac{w_s(\omega)}{\hbar\omega}
\sum_{c\in\{\rm dir,hc\}}Y_{c,r}(\omega,s)P_c^{(s)}(\omega)\dd\omega.
\label{eq:Susefulgeneration}
\end{equation}
Here, $w_s(\omega)\geq0$ is a dimensionless quadrature weight over the discrete direction/polarization scenarios, $\hbar$ is the reduced Planck constant, and $G_r[\mathsf T]$ is the areal useful-event generation rate in m$^{-2}$~s$^{-1}$. The channel sum is restricted by Eq.~\eqref{eq:Spartition}; unused energy enters Eq.~\eqref{eq:Sheat}.

\begin{table}[H]
\centering
\caption{Parameters of the finite-dimensional matrix-oscillator demonstration. These values define a conditional finite model, not a universal material constant.}
\label{tab:matrixparams}
\begin{tabular}{lr}
\toprule
Parameter & Value \\
\midrule
Equivalent optical thickness & 5.0 nm \\
Plasma energy & 9.01 eV \\
Static response eigenvalue & 3.0 \\
Chemical load & 1.23 eV \\
Direct-channel center / width & 2.20 / 0.35 eV \\
Hot-carrier barrier & 0.80 eV \\
Ambient temperature & 300 K \\
Thermal coefficient / emissivity & 20.0 W m$^{-2}$ K$^{-1}$ / 0.9 \\
\bottomrule
\end{tabular}
\end{table}

The finite two-dimensional oscillator program yields a conditional hybrid limit of 4.3096\% for frequency-dependent positive-semidefinite atoms, 4.2820\% for one common two-axis eigenbasis, and 2.5788\% for one polarization axis. These values diagnose the consequence of progressively restricting the response space. They are not material-specific efficiencies, because the finite oscillator budgets and channel kernels in \supptab~\ref{tab:matrixparams} are conditional.

\subsection*{4.3 Quasistatic material screen}

For a spheroid of volume $V$ embedded in a host with relative permittivity $\epsilon_{\rm b}$, define the relative contrast susceptibility $\chi=\epsilon/\epsilon_{\rm b}-1$ and principal-axis depolarization factors $L_i$, satisfying $\sum_iL_i=1$. The orientation-averaged extinction cross-section $\sigma_{\rm ext}$ obeys
\begin{equation}
\frac{\sigma_{\rm ext}}{V}=\frac{2\pi}{3\lambda}
\sum_{i=1}^{3}\operatorname{Im}\left[\frac{\chi}{1+L_i\chi}\right].
\label{eq:Sspheroid}
\end{equation}
Here, $\lambda$ is vacuum wavelength and $\operatorname{Im}$ denotes the imaginary part. A sphere has $L_i=1/3$. For silver (Ag), gold (Au) and rhodium (Rh), the material-only limit is the exact finite-loss, two-eigenvalue sum-rule solution associated with the quasistatic bound of Miller \etal\cite{Miller2014}; the low-loss asymptote is retained only as a diagnostic. For equivalent metal thickness $t_m$, an optimistic independent-particle mapping defines a dimensionless screened absorptance
\begin{equation}
A_{\rm screen}(\lambda)=1-
\exp\left[-\frac{\sigma_{\rm ext}(\lambda)}{V}t_m\right].
\label{eq:Sscreen}
\end{equation}
Equation~\eqref{eq:Sscreen} treats extinction as potentially useful absorption and neglects radiative escape, particle coupling, and surface-coverage constraints. Therefore, it is an outer screen.

\begin{table}[H]
\centering
\caption{Optimistic quasistatic material screening at a 1.0-eV Fowler threshold and 5-nm equivalent metal thickness. Extinction is treated as potentially useful absorption.}
\label{tab:materialscreen}
\begin{tabular}{lrrrrr}
\toprule
Metal & Sphere (\%) & Fixed spheroid (\%) & Pointwise spheroid (\%) & Material limit (\%) & Fixed/limit \\
\midrule
Ag & 0.354 & 2.308 & 6.304 & 6.374 & 32.7\% \\
Au & 0.517 & 2.556 & 4.075 & 4.231 & 54.0\% \\
Rh & 0.097 & 5.232 & 5.566 & 5.618 & 91.9\% \\
\bottomrule
\end{tabular}
\end{table}

\begin{figure}[H]
\centering
\includegraphics[width=0.98\textwidth]{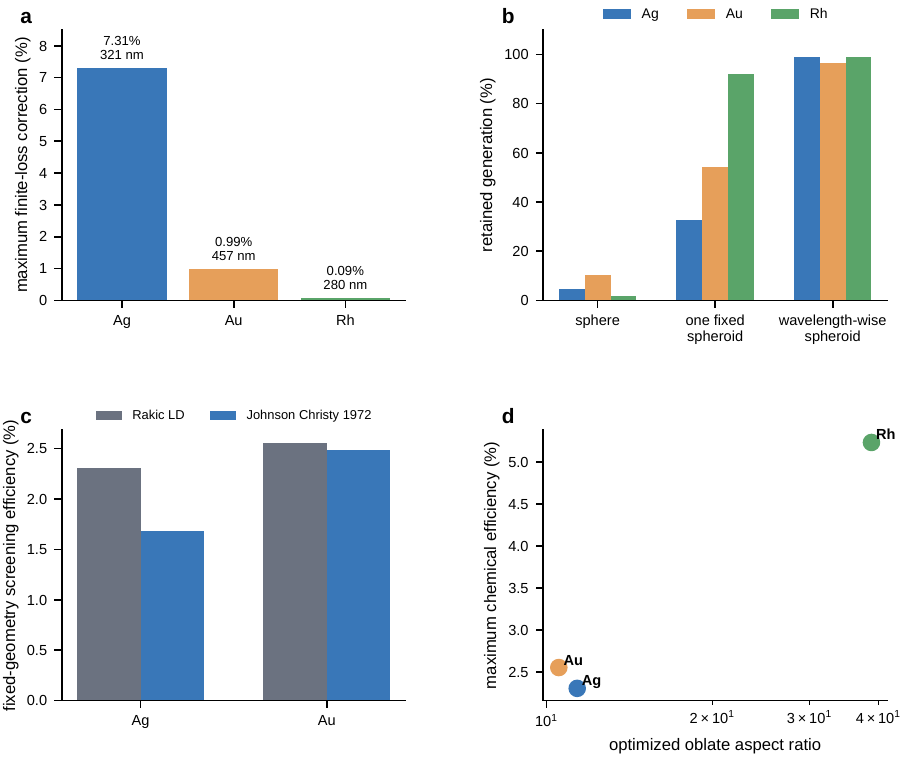}
\caption{\textbf{Quasistatic material and geometry diagnostics.} \textbf{a}, Maximum relative correction of the exact finite-loss material bound to its low-loss asymptote over the simulated spectrum for silver (Ag), gold (Au), and rhodium (Rh). The correction is largest for Ag at short wavelength. \textbf{b}, Useful-generation response of a sphere, one fixed spheroid, and a wavelength-wise optimized spheroid, normalized to the exact material limit. The wavelength-wise envelope changes shape with wavelength and is not one realizable particle. \textbf{c}, Sensitivity of the optimized fixed-geometry screen to alternative optical-constant data sets for Ag and Au. \textbf{d}, Optimum oblate aspect ratio and corresponding fixed-geometry screening efficiency for each metal. All four panels use an optimistic mapping in which extinction can become useful absorption; they are screening bounds, not device predictions.}
\label{fig:Smaterial}
\end{figure}

Supplementary Figure~\ref{fig:Smaterial}a quantifies, when the finite-loss form is required instead of the low-loss asymptote. Supplementary Figure~\ref{fig:Smaterial}b separates material response from geometry realizability and shows the large loss incurred by one fixed shape for Ag and Au. Supplementary Figure~\ref{fig:Smaterial}c shows that narrow fixed resonances are more sensitive to the chosen optical constants than the broad material limit. Supplementary Figure~\ref{fig:Smaterial}d reports the aspect ratios producing the fixed-shape optima. The numerical values are tabulated in Supplementary Table~\ref{tab:materialscreen}.

The geometry hierarchy is
\begin{equation}
\text{sphere}\subset\text{one fixed spheroid}\subset
\text{wavelength-wise spheroid envelope}\subset\text{material limit}.
\label{eq:Sgeometryhierarchy}
\end{equation}
The wavelength-wise envelope changes geometry with wavelength and is not a realizable single particle. Its proximity to the material limit, contrasted with the much lower fixed-shape retention for Ag and Au, isolates broadband realizability as a leading source of looseness.

\noteheading{5. One-common-structure correlations and certified finite bounds}

Let $s$ index wavelength, polarization, and operating state, let $\mathbf r$ denote position, and let $\rho_s(\mathbf r)$ be the binary or continuous material-distribution field assigned to scenario $s$. Solving an independent optimization for each $s$ permits a different geometry $\rho_s(\mathbf r)$. A physical catalyst instead requires
\begin{equation}
\rho_s(\mathbf r)=\rho(\mathbf r)\quad\text{for all }s.
\label{eq:Sonestructure}
\end{equation}
In polarization-current formulations, Eq.~\eqref{eq:Sonestructure} creates cross-scenario quadratic identities. Lifting products of fields and material variables to a positive-semidefinite Gram matrix gives an outer semidefinite programming (SDP) relaxation. Retaining only diagonal scenario blocks gives a no-cross bound; adding off-diagonal identities removes responses that cannot arise from one common object\cite{Shim2024}.

Two finite benchmarks permit exact enumeration. In the seven-voxel model, all $2^7$ binary structures are evaluated at three frequencies. In the Au/p-GaN benchmark, all $2^9$ structures are evaluated at five spectral bins and two polarizations. For each benchmark, let $\mathcal O$ denote the scalar reaction-weighted electromagnetic objective. The exact shared optimum satisfies
\begin{equation}
\mathcal O_{\rm shared}^{\rm exact}\leq \mathcal O_{\rm all-cross}^{\rm SDP}\leq \mathcal O_{\rm no-cross}^{\rm SDP},
\label{eq:Shierarchy}
\end{equation}
Here, the superscripts identify exact enumeration or a semidefinite-programming (SDP) relaxation, and ``all-cross''/``no-cross'' identify whether off-diagonal scenario identities are imposed. The scenario-wise diagnostic is not necessarily an SDP bound, because it is obtained by independently choosing a physical binary structure for each scenario.

\begin{table}[H]
\centering
\caption{One-common-structure and semidefinite-bound hierarchy for two exactly enumerable finite benchmarks.}
\label{tab:boundhierarchy}
\small
\begin{tabularx}{\textwidth}{>{\raggedright\arraybackslash}p{0.40\textwidth}>{\centering\arraybackslash}p{0.20\textwidth}>{\centering\arraybackslash}X}
\toprule
Quantity & Seven-voxel benchmark & Nine-voxel Au/p-type GaN benchmark \\
\midrule
Independent per-scenario diagnostic & 2.535048 & 10.834809 W m$^{-2}$ \\
No-cross semidefinite-programming upper bound & 2.825493 & 24.887495 W m$^{-2}$ \\
All-cross semidefinite-programming upper bound & 2.231024 & 10.978851 W m$^{-2}$ \\
Exact shared structure & 1.677982 & 8.966666 W m$^{-2}$ \\
One-structure penalty & 33.81\% & 17.24\% \\
Cross-correlation tightening & 21.04\% & 55.89\% \\
\bottomrule
\end{tabularx}
\end{table}

\begin{figure}[H]
\centering
\includegraphics[width=0.98\textwidth]{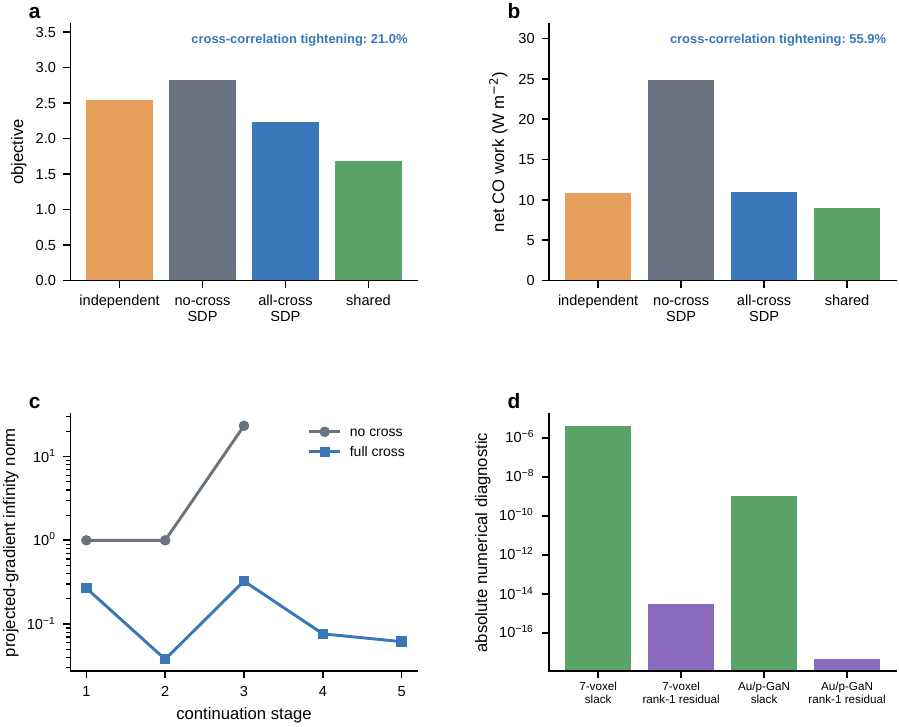}
\caption{\textbf{One-structure certification diagnostics.} \textbf{a}, Objective hierarchy for the exactly enumerable seven-voxel benchmark: independently selected structures, no-cross semidefinite programming (SDP) upper bound, all-cross SDP upper bound, and exact shared structure. \textbf{b}, Corresponding hierarchy for the nine-voxel Au/p-GaN device benchmark. \textbf{c}, Continuation history of the no-cross and full-cross dual searches, plotted as projected-gradient infinity norms. \textbf{d}, Absolute numerical certification diagnostics for the seven-voxel and Au/p-GaN device problems: minimum dual-slack eigenvalues and maximum rank-one cross-identity residuals. Positive slack eigenvalues and near-machine-precision residuals support the stated outer-bound interpretation. Cross correlations substantially tighten the bound, whereas the remaining difference from exact enumeration is a rank-relaxation gap.}
\label{fig:Ssdp}
\end{figure}

Supplementary Figure~\ref{fig:Ssdp}a demonstrates the exact/shared/bound ordering for the abstract three-frequency problem. Supplementary Figure~\ref{fig:Ssdp}b shows the same ordering in the measured-material finite benchmark and the much larger tightening produced by all-cross constraints. Supplementary Figure~\ref{fig:Ssdp}c records the dual continuation and feasibility correction rather than reporting only the final number. Supplementary Figure~\ref{fig:Ssdp}d verifies that physical rank-one structures satisfy the cross identities to numerical precision. Supplementary Table~\ref{tab:boundhierarchy} gives the underlying values. The all-cross bound is rigorous for each declared finite benchmark after the dual slack is shifted by the largest-eigenvalue correction needed to make it positive semidefinite. The correction can loosen the bound but cannot invalidate it. The remaining gap between all-cross SDP and exact enumeration is a rank-relaxation gap. Therefore, the manuscript reports a hierarchy rather than presenting the SDP value as attainable.

The exact 10-bin refinement of the smaller Au/p-GaN device benchmark retains 99.783\% of the independently reoptimized 10-bin shared objective when the 5-bin-selected mask is evaluated at doubled spectral resolution. The 32-voxel constructive design changes by 2.028\% between its 6-bin design grid and 12-bin refinement.

\begin{figure}[H]
\centering
\includegraphics[width=0.92\textwidth]{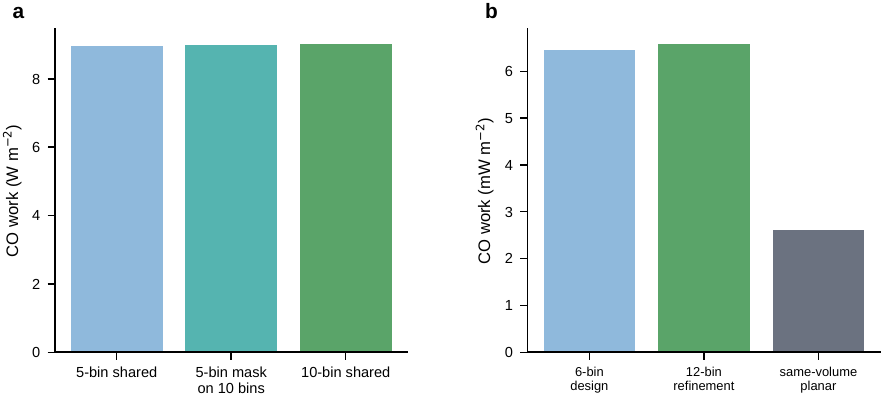}
\caption{\textbf{Independent spectral-refinement checks.} \textbf{a}, Exact shared objective of the nine-voxel benchmark on the five-bin design grid, the five-bin-selected mask reevaluated on ten bins, and the independently reoptimized ten-bin shared objective. Their close agreement tests whether the one-structure hierarchy is a coarse-quadrature artifact. \textbf{b}, Six-bin design objective, twelve-bin reevaluation of the selected 32-voxel source-conditioned structure and same-volume planar comparator. The small design-to-refinement change is contrasted with the much larger three-dimensional-versus-planar difference.}
\label{fig:Sspectral}
\end{figure}

Supplementary Figure~\ref{fig:Sspectral}a shows 99.783\% retention of the independently reoptimized ten-bin objective by the five-bin-selected shared mask. Supplementary Figure~\ref{fig:Sspectral}b shows a 2.028\% change between the six-bin design value and twelve-bin refinement of the larger constructive design, while the shared three-dimensional response remains well above the planar comparator. These two tests independently support the spectral convergence statements in main-text Figures~3 and~5.

\noteheading{6. Exact multi-electron accumulation and low-flux scaling}

Many useful reactions require $n>1$ sequential redox equivalents at the same accumulation reservoir. Consider states $i=0,1,\ldots,n-1$, where $i$ carriers have been retained. Useful arrivals occur at rate $g$. Every incomplete state $i>0$ resets to state 0 at rate $k=1/\tau$. An arrival from state $n-1$ completes product formation and resets the reservoir.

For $1\leq i\leq n-1$, the stationary master equation gives
\begin{equation}
(g+k)p_i=gp_{i-1},
\label{eq:Srecurrence}
\end{equation}
so
\begin{equation}
p_i=\zeta^ip_0,
\qquad
\zeta=\frac{g}{g+k}=\frac{g\tau}{1+g\tau}.
\label{eq:Sr}
\end{equation}
where $\zeta$ is the ratio of successive stationary populations. Normalization, $\sum_{i=0}^{n-1}p_i=1$, yields
\begin{equation}
p_0=\frac{1-\zeta}{1-\zeta^n},
\qquad
p_i=\frac{(1-\zeta)\zeta^i}{1-\zeta^n}.
\label{eq:Spopulations}
\end{equation}
The product-completion rate and carrier utilization are
\begin{equation}
J_n=gp_{n-1}=g\frac{(1-\zeta)\zeta^{n-1}}{1-\zeta^n},
\label{eq:SJn}
\end{equation}
\begin{equation}
U_n=\frac{nJ_n}{g}=n\frac{(1-\zeta)\zeta^{n-1}}{1-\zeta^n}.
\label{eq:SUn}
\end{equation}
For $g\tau\ll1$, $\zeta\simeq g\tau$, hence
\begin{equation}
J_n\simeq g^n\tau^{n-1}.
\label{eq:Slowflux}
\end{equation}
This strong low-flux scaling is the origin of the one-sun multi-electron penalty. It is not captured by raising a single-photon yield to the $n$th power because the waiting-time and reset process is explicit.

\begin{table}[H]
\centering
\caption{Dimensionless arrival--lifetime products required for specified carrier utilization in the exact sequential-reset network.}
\label{tab:gtau}
\begin{tabular}{rrr}
\toprule
Required carriers $n$ & $g\tau$ for 50\% utilization & $g\tau$ for 90\% utilization \\
\midrule
2 & 0.5000 & 4.5000 \\
4 & 2.0479 & 14.1474 \\
6 & 3.6299 & 23.8009 \\
8 & 5.2174 & 33.4552 \\
\bottomrule
\end{tabular}
\end{table}

\begin{figure}[H]
\centering
\includegraphics[width=0.98\textwidth]{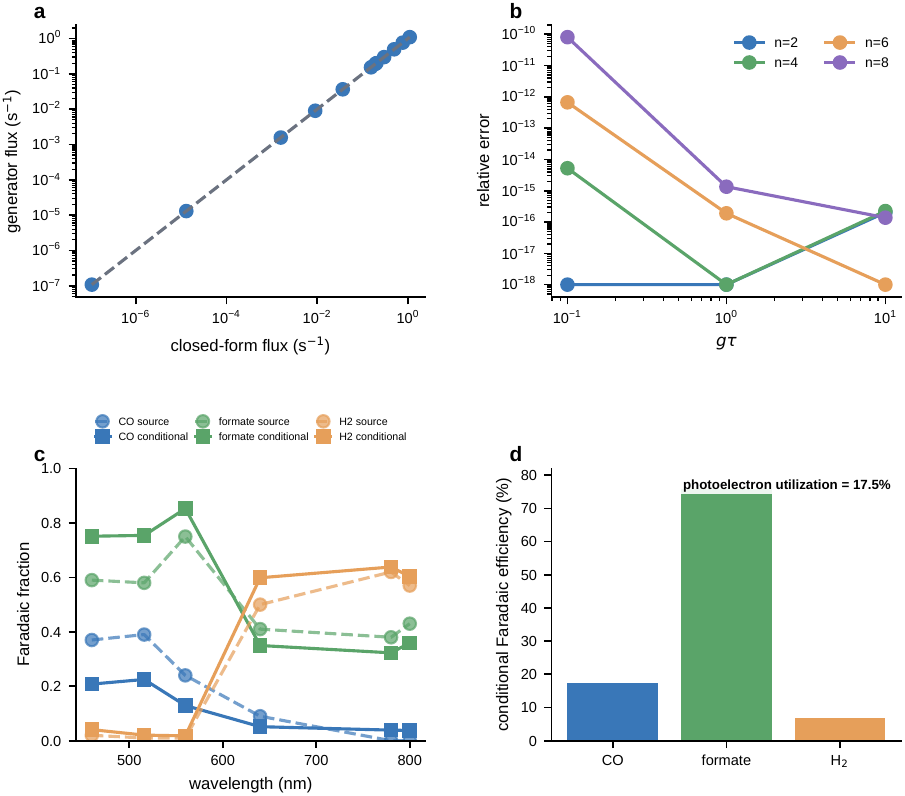}
\caption{\textbf{Kinetic validation and conditional network diagnostics.} \textbf{a}, Product-completion flux from the closed-form sequential-reset expression in Eq.~\eqref{eq:SJn} compared with an independently assembled continuous-time Markov generator. The dashed diagonal denotes exact agreement over the full validation range. \textbf{b}, Relative difference between the two calculations versus the dimensionless arrival--storage product $g\tau$ for products requiring two, four, six, and eight carriers; the maximum discrepancy is $8.11\times10^{-11}$. \textbf{c}, Wavelength-resolved product fractions supplied to the conditional network: carbon monoxide (CO), formate and molecular hydrogen (H$_2$). Dashed curves are source-resolved targets and solid curves are the conditional network outputs, showing that the illustrative network does not exactly fit every source point. \textbf{d}, Conditional one-sun Faradaic product partition for CO, formate, and H$_2$; the annotation gives the fraction of delivered photoelectrons that complete a product cycle. Panels \textbf{c} and \textbf{d} diagnose one admissible source-conditioned kinetic member and are not universal catalyst predictions.}
\label{fig:Skinetics}
\end{figure}

Supplementary Figure~\ref{fig:Skinetics}a compares the closed-form completion flux with an independently assembled Markov generator, while Supplementary Figure~\ref{fig:Skinetics}b resolves their relative error over electron count and $g\tau$. Supplementary Figure~\ref{fig:Skinetics}c shows the wavelength-dependent product partition supplied to the conditional network, and Supplementary Figure~\ref{fig:Skinetics}d shows the resulting conditional Faradaic product distribution. The first two panels validate the universal network; the latter two are explicitly source-conditioned diagnostics.

The maximum relative difference between Eq.~\eqref{eq:SJn} and the independent Markov-generator solution over the validation grid is $8.11\times10^{-11}$. The universal variable is $g\tau$; mapping it to an absolute lifetime requires an explicit site or reservoir density and a delivered carrier flux. Consequently, absolute lifetimes from a conditional geometry are not universal bounds.

\noteheading{7. Source-conditioned Au on p-type GaN device and finite three-dimensional design}

\subsection*{7.1 Data boundary}

The source-conditioned operator uses publicly reported wavelength-dependent product partitions and consumed-charge trends for a Au/p-GaN photocathode under controlled absorbed power\cite{Kiani2026}. The public record supports a wavelength-dependent preference for carbon monoxide (CO) at visible interband excitation and molecular hydrogen (H$_2$) at redder intraband excitation, as well as morphology-dependent hot-carrier transport. It does not provide state-by-state density functional theory (DFT) or many-body Green's-function and screened-Coulomb-interaction ($GW$) optical transition matrices, raw time-resolved partial-current traces or acquisition-duration metadata sufficient to identify absolute $j_p(\lambda)/P_{\rm abs}(\lambda)$.

Accordingly, the product operator is represented as a positive spectral probability
\begin{equation}
\eta_p(\lambda)=\eta_{\rm pk}\,r_Q(\lambda)f_p^{\rm FE}(\lambda),
\label{eq:Ssourceoperator}
\end{equation}
where $p$ indexes CO, formate, or H$_2$, $f_p^{\rm FE}(\lambda)$ is the digitized wavelength-dependent Faradaic product fraction, $r_Q(\lambda)$ is the relative total consumed-charge-per-absorbed-power envelope, $\eta_{\rm pk}$ is the declared peak carrier-transfer probability, and $\eta_p$ is the resulting carrier probability per absorbed photon. Let $E_\gamma=hc/\lambda$ be photon energy and $\varepsilon_p$ the reversible chemical free energy per transferred electron in the same energy units. The gross and net chemical-work fractions per absorbed photon are
\begin{equation}
w_{p,\rm gross}(\lambda)=\eta_p(\lambda)
\min\left(\frac{\varepsilon_p}{E_\gamma},1\right),
\qquad
w_{p,\rm net}(\lambda)=w_{p,\rm gross}(\lambda)-w_{p,\rm rev}(\lambda),
\label{eq:Sproductweight}
\end{equation}
where $w_{p,\rm rev}\geq0$ is the reciprocal ambient-radiation work fraction evaluated with the same product operator and chemical affinity. The residual solar fraction enters heat exactly once:
\begin{equation}
w_{\rm heat}(\lambda)=1-\sum_p w_{p,\rm gross}(\lambda).
\label{eq:Sheatfraction}
\end{equation}
Admissibility requires $0\leq\sum_p w_{p,\rm gross}(\lambda)\leq1$, so $w_{\rm heat}\geq0$ and the gross chemical and heat fractions close exactly. No missing raw transition rows are replaced by synthetic electronic states in the reported source-conditioned result.

\begin{table}[H]
\centering
\caption{Data boundary and finite optical-design results for the Au/p-GaN illustration.}
\label{tab:casescope}
\begin{tabularx}{\textwidth}{>{\raggedright\arraybackslash}p{0.42\textwidth}X}
\toprule
Item & Status or result \\
\midrule
Absolute wavelength-resolved partial-current responsivity identifiable & No \\
Channel operator & Public-figure-resolved positive operator with declared peak-transfer scale \\
Design domain & 4$\times$4$\times$2 binary Au voxels; 70$\times$70$\times$14 nm$^3$; 160-nm pitch \\
Shared three-dimensional carbon-monoxide work, refined grid & 0.006576 W m$^{-2}$ \\
Same-volume planar carbon-monoxide work & 0.002614 W m$^{-2}$ \\
Three-dimensional/planar ratio & 2.466 \\
Scenario-specific heuristic / shared penalty & 0.006904 W m$^{-2}$ / 6.65\% \\
Interpretation & Constructive lower bound for the stated finite design space, not a global material optimum. \\
\bottomrule
\end{tabularx}
\end{table}

\subsection*{7.2 Finite optical model}

The larger constructive design contains $4\times4\times2=32$ binary Au voxels of size $17.5\times17.5\times7$~nm$^3$ in a $70\times70\times14$~nm$^3$ region with 160-nm normalization pitch. Measured Au optical constants and a p-type GaN optical response are used in a retarded volume-integral model with a planar reflected Green tensor. A multistart constructive search samples 12,779 candidate masks subject to a 24-voxel occupancy limit. Because the search is not exhaustive over $2^{32}$ structures, its best design is a lower bound on the finite design-space optimum.

The refined shared CO work is $6.576\times10^{-3}$~W~m$^{-2}$, compared with $2.614\times10^{-3}$~W~m$^{-2}$ for a same-volume planar comparator. The scenario-wise heuristic is $6.904\times10^{-3}$~W~m$^{-2}$, corresponding to a 6.65\% one-structure penalty. The design reaches 10.27\% of a loose unit-absorptance channel bound. None of these values is an asserted solar-to-CO efficiency of the experimental system.

\begin{figure}[H]
\centering
\includegraphics[width=0.98\textwidth]{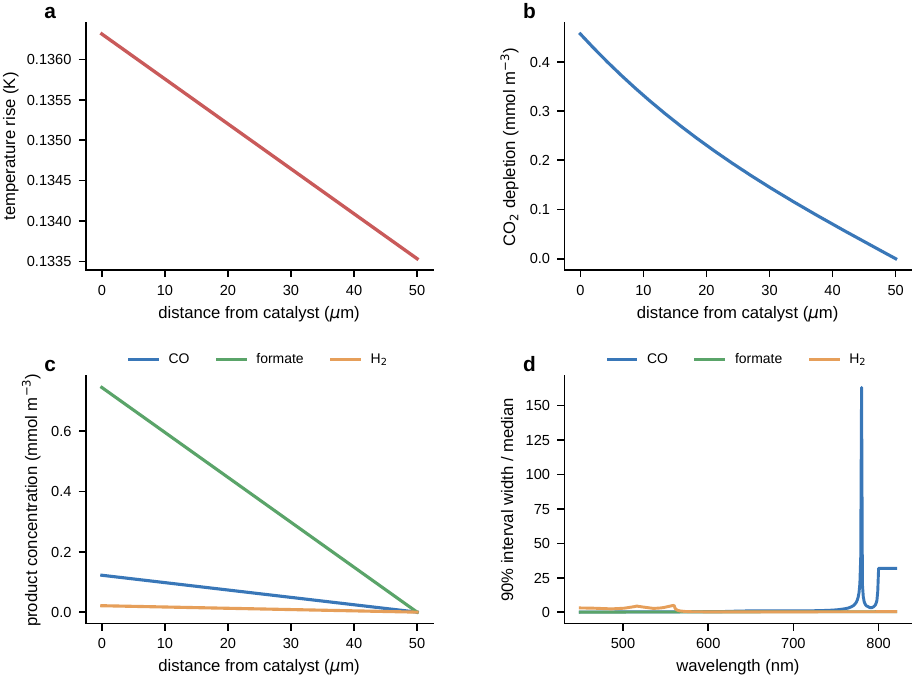}
\caption{\textbf{Continuum-transport and source-operator uncertainty diagnostics.} \textbf{a}, Temperature rise across the one-dimensional electrolyte boundary layer at the conditional one-sun operating point. \textbf{b}, Carbon-dioxide depletion relative to the bulk concentration, expressed in millimoles per cubic metre; the small surface depletion confirms that this conditional point is far below the diffusion limit. \textbf{c}, Carbon-monoxide, formate, and hydrogen concentration profiles generated by the conditional product fluxes. \textbf{d}, Relative width of the propagated 90\% source-operator interval, normalized by the median channel probability. The sharp long-wavelength carbon-monoxide feature reflects division by a median probability approaching zero and is retained rather than clipped.}
\label{fig:Ssource}
\end{figure}

Supplementary Figure~\ref{fig:Ssource}a shows the temperature profile used to close the conditional heat-transfer problem. Supplementary Figure~\ref{fig:Ssource}b displays the corresponding carbon-dioxide depletion, Supplementary Figure~\ref{fig:Ssource}c gives the product concentration profiles and Supplementary Figure~\ref{fig:Ssource}d quantifies the wavelength-dependent uncertainty inherited by the positive channel operator.

The substrate contribution is a leading-order term in this finite model. Removing the reflected Green contribution or replacing the Weyl tensor by a retarded image approximation changes the objective more than spectral refinement. The source-conditioned geometry should therefore be interpreted as a demonstration of the theory's data interface, not as a robust fabrication prescription.

\noteheading{8. Locally detailed-balanced admissible microkinetic family}

The main theoretical result does not require a unique transition-state landscape. To illustrate how finite kinetics can be added without violating thermodynamics, a closed six-state network is used:
\begin{equation}
*,\quad \mathrm{CO_2^*},\quad \mathrm{COOH^*},\quad \mathrm{CO^*},\quad \mathrm{OCHO^*},\quad \mathrm{H^*}.
\label{eq:Sstates}
\end{equation}
Here, $*$ denotes a vacant catalytic site; $\mathrm{CO_2^*}$, $\mathrm{COOH^*}$, $\mathrm{CO^*}$ and $\mathrm{OCHO^*}$ denote adsorbed carbon dioxide, carboxyl, carbon monoxide, and formate intermediates, respectively; and $\mathrm{H^*}$ denotes adsorbed hydrogen. Reversible dark steps comprise CO$_2$ adsorption, two proton-coupled electron-transfer steps to CO, CO desorption, two steps to formate, and two hydrogen-evolution steps. Source-conditioned photo-steps are added in parallel. Each dark pair satisfies Eq.~\eqref{eq:Sldb}; therefore, fitting forward scales or barrier offsets cannot independently alter the equilibrium constant.

More generally, an admissible kinetic family may be written
\begin{equation}
\mathcal K=\left\{\kappa_\nu^\pm:\
\ln\frac{\kappa_\nu^+}{\kappa_\nu^-}=-\frac{\Delta G_\nu}{\kB T},\quad
0\leq \kappa_\nu^\pm\leq \kappa_{\nu,\max}\right\}.
\label{eq:SadmissibleK}
\end{equation}
Here, $\nu$ indexes an elementary reversible reaction step, $\kappa_\nu^\pm$ are its forward/reverse rate constants, $\Delta G_\nu=G_{\nu,\rm products}-G_{\nu,\rm reactants}$ is the forward reaction free-energy change per event at the declared potential, composition and temperature, and $\kappa_{\nu,\max}$ is a declared kinetic upper bound. Let $\boldsymbol\kappa$ collect all elementary forward and reverse rates. Product-flux bounds then follow from
\begin{equation}
\mathcal J_r^{\max}=\sup_{\boldsymbol\kappa\in\mathcal K}\mathcal J_r,
\qquad
\mathcal J_r^{\min}=\inf_{\boldsymbol\kappa\in\mathcal K}\mathcal J_r.
\label{eq:Skineticbounds}
\end{equation}
The fitted six-state point used for the energy-ledger illustration is one member of such a family, not a first-principles transition-state prediction. Its local-detailed-balance and stationarity residuals are numerical consistency checks, while its product selectivity is explicitly conditional.

\begin{table}[H]
\centering
\caption{Conditional one-sun microkinetic and transport operating point. The fitted offsets are source-conditioned sensitivity parameters rather than first-principles transition-state free energies.}
\label{tab:conditional}
\small
\begin{tabularx}{\textwidth}{>{\raggedright\arraybackslash}p{0.54\textwidth}X}
\toprule
Quantity & Value \\
\midrule
Photoelectron utilization & 17.54\% \\
Faradaic efficiency: carbon monoxide (CO) & 17.44\% \\
Faradaic efficiency: formate & 75.83\% \\
Faradaic efficiency: molecular hydrogen (H$_2$) & 6.73\% \\
CO product flux & 4.964e-09 mol m$^{-2}$ s$^{-1}$ \\
Formate product flux & 2.159e-08 mol m$^{-2}$ s$^{-1}$ \\
H$_2$ product flux & 1.917e-09 mol m$^{-2}$ s$^{-1}$ \\
Local detailed-balance residual & 7.105e-15 \\
Stationarity residual & 3.469e-18 s$^{-1}$ \\
Surface temperature and hydrogen-ion activity (pH) & 298.286 K; 6.80001 \\
Carbon-dioxide transport-limit fraction & 1.383e-05 \\
Selectivity-fit residual norm & 0.401 \\
\bottomrule
\end{tabularx}
\end{table}

At this point, only 17.54\% of delivered photoelectrons complete a product cycle. This result emphasizes why an optical carrier-generation limit cannot be identified with chemical efficiency. The exact value depends on the illustrative network and should not be transferred to other catalysts.

\noteheading{9. Continuum heat and mass transfer}

A one-dimensional stagnant boundary layer of thickness $L$ is used as a transport closure, with coordinate $x=0$ at the catalyst and $x=L$ in the well-mixed bulk. For neutral species $i$, let $c_i(x)$ be molar concentration, $D_i$ its diffusion coefficient, and $\mathcal N_i$ its positive-outward molar flux. Then
\begin{equation}
\mathcal N_i=-D_i\frac{\dd c_i}{\dd x},
\qquad
\frac{\dd \mathcal N_i}{\dd x}=0.
\label{eq:Sdiffusion}
\end{equation}
At $x=0$, $\mathcal N_i$ equals the microkinetic consumption or production rate; at $x=L$, $c_i=c_{i,\rm bulk}$. Heat transport uses temperature $T(x)$ and liquid thermal conductivity $k_{\rm th}$ and satisfies
\begin{equation}
-\frac{\dd}{\dd x}\left(k_{\rm th}\frac{\dd T}{\dd x}\right)=0
\label{eq:Sheateq}
\end{equation}
within the liquid film, with residual absorbed power as a surface source and an external heat-transfer boundary condition at $x=L$. This closure excludes three-dimensional convection around the photo-scanning electrochemical microscopy (photo-SECM) tip and is used only to compare the ordering of heating and mass-transfer onset.

At one sun, the conditional surface-temperature rise is 0.136~K and the carbon flux is $1.383\times10^{-5}$ of the estimated CO$_2$ diffusion limit. At $10^3$ concentration, the model gives a temperature rise near 136~K, while the transport-limit fraction remains below $10^{-2}$. Thus, under the stated parameters, heat removal becomes important before CO$_2$ depletion. The high-concentration temperature should not be interpreted literally, once temperature-dependent material properties, convection, or boiling become relevant.

\begin{figure}[H]
\centering
\includegraphics[width=0.94\textwidth]{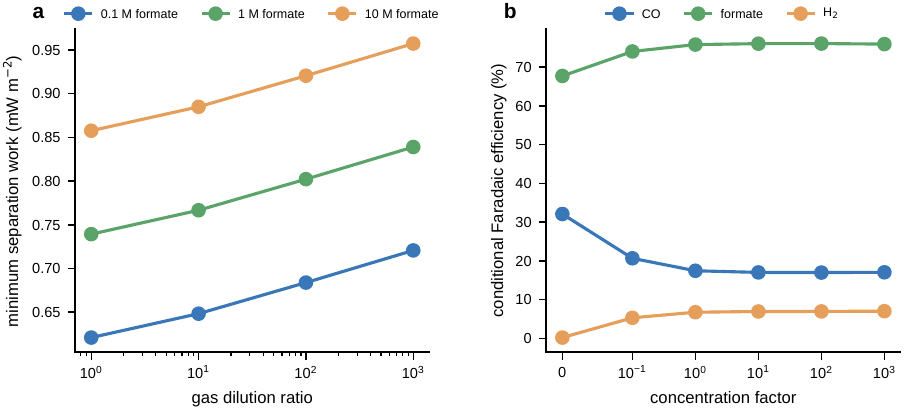}
\caption{\textbf{Separation and optical-concentration sensitivities.} \textbf{a}, Ideal minimum reversible separation work versus gas-product dilution for three target formate concentrations. The gas and dissolved-product contributions are evaluated from the mixing free energies in Eqs.~\eqref{eq:Sgassep} and \eqref{eq:Ssolsep}. \textbf{b}, Conditional carbon-monoxide, formate, and hydrogen Faradaic efficiencies versus optical concentration within the six-state kinetic model and one-dimensional heat/mass-transfer closure. The curves show selectivity redistribution within the conditional model and do not represent measured concentration-dependent efficiencies.}
\label{fig:Stransportsep}
\end{figure}

Supplementary Figure~\ref{fig:Stransportsep}a separates the thermodynamic cost of gas dilution from the cost of concentrating dissolved formate. Supplementary Figure~\ref{fig:Stransportsep}b shows how the illustrative reaction network redistributes product selectivity as optical concentration changes. Together with the profiles in Supplementary Figure~\ref{fig:Ssource}c, these panels support the statement that heat removal becomes relevant before CO$_2$ depletion under the declared transport closure.

\noteheading{10. Applied-bias and product-separation ledger}

For product $p$, let $\mathcal J_p>0$ denote net production as an areal molar flux, $n_p$ its electron stoichiometry, $F$ Faraday's constant, and $V_p^{\rm rev}$ its reversible electrochemical potential in volts. Chemical power density is
\begin{equation}
P_{\rm chemical}=\sum_p n_pF V_p^{\rm rev}\mathcal J_p.
\label{eq:Schemicalpower}
\end{equation}
For gas-product set $\mathcal P_{\rm gas}$ with outlet mole fractions $y_p$, the ideal reversible separation work is
\begin{equation}
P_{\rm sep,gas}=RT\sum_{p\in\mathcal P_{\rm gas}}\mathcal J_p\ln\frac{1}{y_p}.
\label{eq:Sgassep}
\end{equation}
For dissolved-product set $\mathcal P_{\rm sol}$, with feed and target concentrations $c_{p,\rm feed}$ and $c_{p,\rm target}$,
\begin{equation}
P_{\rm sep,sol}=RT\sum_{p\in\mathcal P_{\rm sol}}\mathcal J_p
\ln\frac{c_{p,\rm target}}{c_{p,\rm feed}}.
\label{eq:Ssolsep}
\end{equation}
Here, $R$ is the molar gas constant and $T$ is the separation temperature. The conservative external electrical work density is
\begin{equation}
P_{\rm bias}=|j_{\rm light}-j_{\rm dark}|V_{\rm cell},
\label{eq:Sbias}
\end{equation}
where $j_{\rm light}$ and $j_{\rm dark}$ are illuminated and dark current densities and $V_{\rm cell}$ is full-cell voltage. With $\Delta$ denoting illuminated-minus-dark increments, the incremental net useful power density is
\begin{equation}
\Delta P_{\rm net}=\Delta P_{\rm chemical}-\Delta P_{\rm sep}-\Delta P_{\rm bias}.
\label{eq:Snet}
\end{equation}
A less conservative attribution that subtracts only current assigned to dark electrolysis is included as sensitivity but is not used for the headline ledger, because a circuit-resolved partition is unavailable.

For a generic delivered chemical-power limit $P_{\rm ch}^0$ (gross product free-energy rate at $U=1$ before separation and bias), productive carrier utilization $U$, separation fraction $f_{\rm sep}$, and incremental external current density $j_{\rm ext}$,
\begin{equation}
P_{\rm net}=UP_{\rm ch}^0(1-f_{\rm sep})-j_{\rm ext}V_{\rm cell},
\label{eq:Sgenericnet}
\end{equation}
so the break-even voltage is
\begin{equation}
V_{\rm BE}=\frac{UP_{\rm ch}^0(1-f_{\rm sep})}{j_{\rm ext}}.
\label{eq:Sbreak}
\end{equation}
Here, $V_{\rm BE}$ is the break-even full-cell voltage at which $P_{\rm net}=0$. This relation is the general result; the following table is a conditional point used to illustrate its application.

\begin{table}[H]
\centering
\caption{Conditional incremental light-minus-dark energy ledger.}
\label{tab:ledger}
\begin{tabular}{lr}
\toprule
Term & Power (W m$^{-2}$) \\
\midrule
Chemical product free energy & +0.007343 \\
Minimum reversible product separation & -0.000802 \\
Externally supplied electrical work & -0.012701 \\
Net useful power & -0.006160 \\
Conditional full-cell voltage & 2.420 V \\
Break-even voltage at fixed kinetics & 1.246 V \\
\bottomrule
\end{tabular}
\end{table}

The electrode potential conversion assumes a specified Ag/Ag-chloride (Ag/AgCl) reference-electrode offset and hydrogen-ion activity (pH) because the full reference-electrode filling solution was not available in the public source. The resulting 2.42-V cell voltage and 1.246-V break-even voltage are therefore sensitivity parameters, not reconstructed device quantities.

\noteheading{11. Numerical implementation and verification}

The numerical analysis was organized into independent modules for the chemical-diode reduction, electromagnetic limits, multi-electron kinetics, the source-conditioned case study, and the system energy ledger. Each reported value was regenerated from the equations, numerical parameters, and source classifications stated in the Main manuscript and Supplementary Information. The modular implementation permits independent evaluation of the analytic reductions, finite electromagnetic benchmarks, kinetic solutions, and energy-accounting closure.

The source audit distinguishes four data classes:
\begin{enumerate}[leftmargin=2em]
\item \textbf{Public primary data}: the Air Mass 1.5 Global solar spectrum tabulated in ASTM International G173 and public optical constants.
\item \textbf{Public-figure-resolved observations}: wavelength-dependent product partitions and relative consumed-charge trends in the Au/p-GaN device illustration.
\item \textbf{Declared conditional parameters}: channel-transfer scale, finite oscillator budgets, kinetic offsets, boundary-layer dimensions, heat-transfer coefficient and reference-electrode conversion.
\item \textbf{Derived quantities}: channel operators, finite optical objectives, carrier utilization, transport profiles, and energy ledgers.
\end{enumerate}

\begin{table}[H]
\centering
\caption{Selected independent numerical verification checks.}
\label{tab:verification}
\begin{tabularx}{\textwidth}{>{\raggedright\arraybackslash}p{0.49\textwidth}X}
\toprule
Check & Result \\
\midrule
Air Mass 1.5 Global power from ASTM International G173 & 1000.3707 W m$^{-2}$ \\
Reduced chemical-diode tests & 6/6 passed \\
Matrix-oscillator and one-structure tests & 10/10 passed \\
Ten-bin mask retention & 99.783\% \\
Local detailed-balance residual & $7.105e-15$ \\
Network-stationarity residual & $3.469e-18\,\mathrm{s}^{-1}$ \\
closed form versus Markov generator & $8.106e-11$ maximum relative error \\
\bottomrule
\end{tabularx}
\end{table}

Numerical checks include the ASTM International G173 one-sun integral; Lambert-$W$ stationarity; containment of spheroidal responses by the finite-loss material limit; exact-enumeration/semidefinite-programming ordering; Markov-generator agreement; spectral refinement; local detailed balance; network stationarity; and arithmetic closure of the energy ledger. Supplementary Table~10 reports the corresponding verification values and tolerances.

\noteheading{12. Proposal for experimental validation}

\textbf{Status and purpose.} No experiment described in this note was performed for the present theoretical study. The protocol defines the measurements and metadata required to calibrate the source-conditioned operator with wavelength-resolved operando data. For every measurement, the instrument manufacturer, model, serial number, acquisition settings, and calibration date should be recorded. The primary outputs are wavelength-resolved absorbed power, absolute product-resolved partial currents, complete electrode and electrolyte metadata, temperature and mass-transfer responses, and a light-minus-dark electrical/separation ledger with propagated uncertainty.

\subsection*{12.1 Preregistered samples, controls, and replication}

The experimental matrix should contain at least four working-electrode classes: bare p-type GaN, a nominally planar Au film with the same total gold loading as the nanostructured samples, a sub-100-nm gold morphology, and a larger morphology centered near 250--350~nm wavelength. The final dimensions should match the structures selected for comparison with the source-conditioned model rather than being inferred after the measurements. A minimum of three independently fabricated batches per class, three working electrodes per batch, and three repeated wavelength cycles per electrode is recommended. Sample order, illumination wavelength, and optical-power sequence should be randomized within each day to separate morphology effects from drift. Dark measurements, CO$_2$-free electrolyte controls and a non-plasmonic metal, or optically inactive control should be included whenever chemically compatible.

For every wafer and electrode, record sapphire orientation and thickness; p-type GaN thickness, magnesium doping, activated hole concentration and mobility; exposed geometric area; ohmic-contact metal sequence, thickness and annealing history; encapsulation geometry; Au mass loading; nanostructure lateral dimensions, height, edge radius, pitch and surface coverage; ligand-removal or plasma-cleaning history; and elapsed time between fabrication and measurement. The fabrication record should specify solvent cleaning, ultraviolet--ozone or oxygen-plasma treatment, contact photolithography or electron-beam lithography, metal deposition rate and base pressure, lift-off or dewetting conditions, annealing ramp and ambient, backside/contact passivation, and electrolyte-window definition. Witness substrates should accompany every deposition and anneal. The analysis should use measured distributions of dimensions and material properties, not only nominal design values.

\subsection*{12.2 Electrochemical cell, electrolyte, and reference calibration}

Use a gas-tight three-electrode photoelectrochemical cell with the Au/p-GaN sample as working electrode, a counter electrode separated by a frit or membrane, and a Ag/AgCl reference electrode whose filling-solution concentration is explicitly reported. The source-centered electrolyte is 0.1~M cesium bicarbonate saturated with CO$_2$. Prepare electrolyte from trace-metal-grade reagents and ultrapure water, record total organic carbon when contamination is a concern, purge for at least 30~min before measurement, and maintain a defined CO$_2$ flow during acquisition. Record cell volume, headspace volume, flow rate, temperature, pressure, initial and final pH, conductivity and dissolved CO$_2$ concentration or the equilibrium calculation used to estimate it.

Calibrate the Ag/AgCl reference against a reversible hydrogen electrode before and after each measurement day. Report the measured offset, filling solution, junction type and drift. Convert potentials using the measured reference offset and measured local/bulk pH; do not rely only on a nominal handbook value. Determine uncompensated resistance by electrochemical impedance spectroscopy at the operating potential and report both uncorrected and resistance-corrected potentials and currents. The source-centered cathodic condition of approximately $-1.5$~V versus Ag/AgCl should be embedded in a potential series, for example from $-1.2$ to $-1.7$~V in 0.1-V increments, to identify whether wavelength selectivity persists when overpotential changes. At each condition, collect dark stabilization, chopped-light transients and steady illumination for a predeclared duration of at least 300~s after the current reaches a stable regime.

\begin{figure}[H]
\centering
\includegraphics[width=0.98\textwidth]{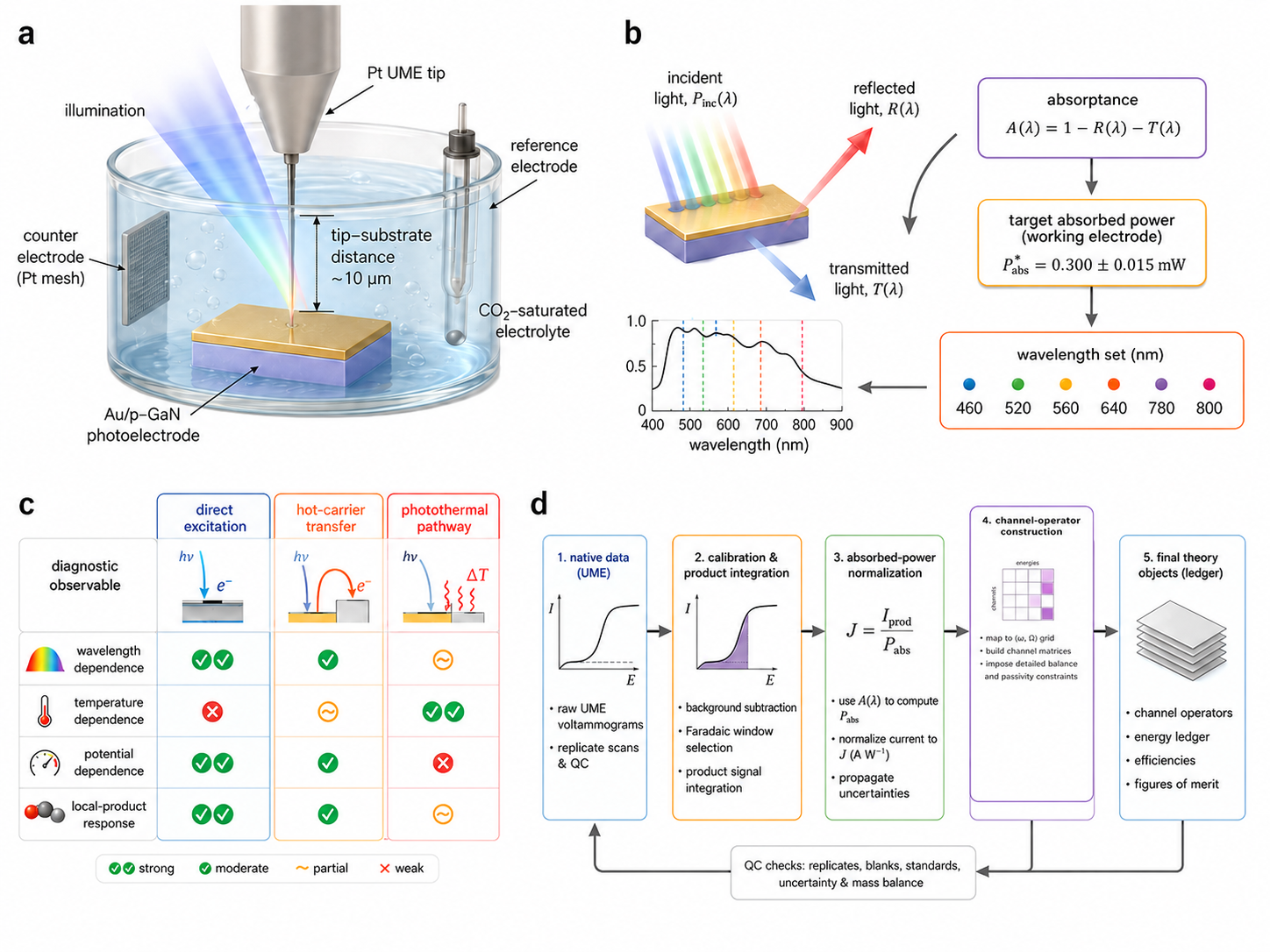}
\caption{\textbf{Experimental validation workflow for an operando-calibrated channel operator.} \textbf{a}, Common photo-scanning electrochemical microscopy geometry for a Au on p-type GaN working electrode. Monochromatic illumination addresses the region directly beneath a platinum ultramicroelectrode tip; the working, counter, and Ag/Ag-chloride reference electrodes are immersed in carbon-dioxide-saturated electrolyte, and the tip--substrate distance is independently calibrated. \textbf{b}, Equal-absorbed-power wavelength protocol. Reflectance $R(\lambda)$ and transmittance $T(\lambda)$ are measured in the electrolyte cell, absorptance is calculated as $A(\lambda)=1-R(\lambda)-T(\lambda)$, and incident power is adjusted to maintain $0.300\pm0.015$~mW absorbed power at the working electrode for the indicated wavelength set. \textbf{c}, Mechanism-discrimination matrix. Wavelength, temperature, potential, and local-product responses provide complementary sensitivity to direct excitation, nonequilibrium hot-carrier transfer, and photothermal conversion. Double green checks, single green checks, amber tildes, and red crosses denote strong, moderate, partial, and weak diagnostic sensitivity, respectively; no single observable uniquely assigns a mechanism. \textbf{d}, Audited data path. Native ultramicroelectrode data pass through background subtraction and product integration, absorbed-power normalization, non-negative channel-operator construction, and the final energy ledger. Quality-control feedback tracks replicates, blanks, standards, uncertainty, and mass balance across all stages.}
\label{fig:Sexperimental}
\end{figure}

Supplementary Figure~\ref{fig:Sexperimental}a fixes a common measurement geometry so that optical, electrochemical and product measurements refer to the same illuminated region. Supplementary Figure~\ref{fig:Sexperimental}b defines wavelength-by-wavelength absorptance and equal-absorbed-power calibration. Supplementary Figure~\ref{fig:Sexperimental}c summarizes orthogonal observables that distinguish direct excitation, hot-carrier transfer, and photothermal conversion. Supplementary Figure~\ref{fig:Sexperimental}d defines the provenance chain from native instrument files to calibrated product currents, channel operators, uncertainty intervals, and the final energy ledger. Supplementary Table~11 specifies the sample, device-stack and physical-characterization record; Supplementary Table~12 specifies optical and thermal calibration; Supplementary Table~13 specifies electrochemical operation, local product detection, and bulk product analysis; and Supplementary Table~14 specifies quality-control, uncertainty, and data-reporting criteria.

\begin{table}[H]
\centering
\caption{Sample, device-stack and physical-characterization record for experimental validation.}
\label{tab:experimental_sample}
\small
\begin{tabularx}{\textwidth}{>{\raggedright\arraybackslash}p{0.23\textwidth}>{\raggedright\arraybackslash}p{0.35\textwidth}X}
\toprule
Category & Required record & Proposed starting condition or acceptance rule \\
\midrule
p-type GaN substrate & Supplier, wafer lot, crystal orientation, layer thickness, magnesium concentration, activated-hole density, mobility, and resistivity & Use one wafer lot for a comparative wavelength series; retain an unpatterned piece as a bare-substrate control. \\
Au nanostructure & Fabrication route, mask or colloid batch, nominal thickness, lateral dimensions, pitch, adhesion layer, annealing, and exposed geometric area & Prepare at least three independent devices per geometry plus same-volume planar-Au and bare-substrate controls. \\
Surface preparation & Solvent, ultraviolet-ozone or plasma treatment, rinse, drying, and elapsed time before measurement & Fix the sequence in advance and keep the delay to electrolyte contact within a declared tolerance. \\
Electron microscopy & Scanning electron microscope; report manufacturer, model, accelerating voltage, working distance, detector and image scale & Acquire at least five fields of view per device and report dimension distributions rather than a single representative image. \\
Topography & Atomic force microscope; report manufacturer, model, mode, tip, scan size and line rate & Report root-mean-square roughness and height distributions for patterned, planar and bare controls. \\
Optical constants and thickness & Spectroscopic ellipsometer; report manufacturer, model, angular and spectral range, and fitting model & Fit gold, p-type gallium nitride and any adhesion layer using the same dispersion model employed in electrodynamic calculations. \\
Chemical/electronic state & X-ray photoelectron and ultraviolet photoelectron spectrometers; report manufacturer, model and acquisition conditions & Record before and after electrolysis; report binding-energy calibration, work function and detectable surface contamination. \\
Electrical contact & Metallization sequence, annealing, contact area, sheet/contact resistance and encapsulation geometry & Confirm ohmic behaviour over the current range used for illumination experiments. \\
\bottomrule
\end{tabularx}
\end{table}

\begin{table}[H]
\centering
\caption{Optical and thermal calibration protocol. Ultraviolet-visible-near-infrared measurements, equal-absorbed-power exposure and local thermometry are specified together with the instrument metadata required for reproducibility.}
\label{tab:experimental_optical}
\small
\begin{tabularx}{\textwidth}{>{\raggedright\arraybackslash}p{0.23\textwidth}>{\raggedright\arraybackslash}p{0.37\textwidth}X}
\toprule
Measurement & Required implementation details & Proposed acceptance rule \\
\midrule
Reflectance and transmittance & Ultraviolet-visible-near-infrared spectrophotometer with integrating sphere; report manufacturer and model; reference standards, incidence angle, polarization and electrolyte optical path & Determine $A(\lambda)=1-R(\lambda)-T(\lambda)$ on the same wetted device state used for chemistry; energy closure $|1-R-T-A_{\rm parasitic}|\leq0.03$. \\
Monochromatic illumination & Tunable laser, supercontinuum source plus monochromator, or discrete lasers; report manufacturer and model; linewidth, polarization, beam diameter and spatial profile & Cover the product-selective range with predeclared wavelengths; verify wavelength at the sample plane and avoid changing the illuminated area between wavelengths. \\
Power calibration & National-metrology-institute-traceable photodiode or power meter; report manufacturer and model positioned at the sample plane & Maintain incident-power uncertainty below 2\% and absorbed-power mismatch between wavelengths below 3\%. \\
Equal-absorbed-power scan & Compute $P_{\rm inc}(\lambda)=P_{\rm abs,target}/A(\lambda)$ and verify after every optical realignment & Randomize wavelength order; repeat the first wavelength at the end and require drift below 5\%. \\
Beam profile & Camera or knife-edge beam profiler; report manufacturer and model & Report $1/e^2$ diameter, ellipticity and overlap with the electrochemically active area. \\
Surface temperature & Calibrated infrared microscope, micro-Raman thermometer or embedded microthermocouple; report manufacturer, model and calibration & Establish emissivity/temperature calibration on a dark-heated reference; report temporal response and uncertainty, preferably $\leq0.2$ K near one sun. \\
Modulated response & Optical chopper and lock-in amplifier or digitally modulated source; report manufacturer and model & Measure amplitude and phase over at least two decades of modulation frequency to separate prompt electronic and slower thermal responses. \\
Temperature-matched control & Resistive heater with closed-loop controller; report manufacturer and model & Reproduce the illuminated surface temperature in the dark within the thermometry uncertainty while keeping electrochemical potential and mass transport unchanged. \\
\bottomrule
\end{tabularx}
\end{table}

\begin{table}[H]
\centering
\caption{Electrochemical, product-analysis and operando scanning electrochemical microscopy protocol.}
\label{tab:experimental_electrochem}
\small
\begin{tabularx}{\textwidth}{>{\raggedright\arraybackslash}p{0.22\textwidth}>{\raggedright\arraybackslash}p{0.39\textwidth}X}
\toprule
Component & Required record & Proposed starting condition or validation rule \\
\midrule
Cell and potentiostat & Gas-tight three-electrode cell; potentiostat with manufacturer and model reported; working, counter and reference positions; exposed area; cell volume & Use the identical cell geometry for all wavelengths and dark controls; record uncompensated resistance before and after each run. \\
Electrolyte & Reagent grade, concentration, volume, dissolved carbon-dioxide protocol, measured pH, conductivity, temperature and gas flow & A source-matched starting point is 0.1 M cesium bicarbonate saturated with carbon dioxide at $298.15\pm0.5$ K; verify rather than assume the final pH and dissolved-gas state. \\
Reference electrode & Silver/silver-chloride or other reference electrode with manufacturer and model reported, filling solution, junction and placement & Calibrate against the reversible hydrogen electrode before and after the campaign; reference drift should be below 5 mV. \\
Counter electrode & Material, area, compartment/separator and distance from working electrode & Confirm that counter products do not reach the analytical volume or the scanning probe. \\
Potential program & Chronoamperometry, potential steps or cyclic voltammetry; scan rate, hold time, preconditioning and dark/light sequence & Use randomized light/dark blocks with at least three independent devices; require a declared steady-state criterion before product integration. \\
Ohmic correction & Electrochemical impedance spectroscopy; report potentiostat manufacturer and model, frequency/amplitude range and fitted circuit & Report raw and corrected potentials; do not apply undocumented post hoc compensation. \\
Gas products & Gas chromatograph with manufacturer and model reported with thermal-conductivity and/or flame-ionization detection and methanizer as required & Use at least five calibration levels bracketing the sample; include blank, carry-over and standard-recovery checks. \\
Dissolved formate & Nuclear magnetic resonance spectrometer, ion chromatograph or high-performance liquid chromatograph with manufacturer and model reported and internal standard & Report calibration curve, limit of detection, sample volume, dilution and recovery; analyse post-electrolysis blanks. \\
Partial current & Convert product rate to $j_p=n_pF\dot N_p/A_{\rm geo}$ and normalize by measured absorbed power & Report both geometric-area and electrochemically active-area normalizations when available; propagate optical and analytical uncertainty. \\
Operando scanning electrochemical microscopy & Platinum ultramicroelectrode with manufacturer and model reported, tip radius, approach curve, tip--sample distance, voltammetric window, scan rate and product calibration & A source-matched geometry may begin near a 1~$\mu$m tip radius and 10~$\mu$m gap, but both quantities must be measured for each experiment. \\
\bottomrule
\end{tabularx}
\end{table}

\begin{table}[H]
\centering
\caption{Quality-control, uncertainty and data-reporting requirements for an operando-calibrated channel operator.}
\label{tab:experimental_acceptance}
\small
\begin{tabularx}{\textwidth}{>{\raggedright\arraybackslash}p{0.30\textwidth}X}
\toprule
Item & Required criterion or reported data \\
\midrule
Independent replication & At least three independently fabricated devices per geometry and three repeated wavelength sequences per device; report all exclusions. \\
Absorbed-power consistency & Relative mismatch below 3\% across wavelength conditions; provide raw incident-power, reflectance, transmittance and beam-profile files. \\
Current stability & Predeclared steady-state criterion, for example slope below 1\% of the mean current per minute over the final analysis window. \\
Product mass and charge balance & Sum of quantified Faradaic efficiencies between 90 and 110\% unless an independently identified product explains the deviation. \\
Blank controls & Dark, bare-substrate, planar-gold, no-carbon-dioxide and temperature-matched dark controls acquired with the same timing and potential program. \\
Reference and resistance stability & Reference drift below 5 mV and uncompensated-resistance change below 10\% during a wavelength series. \\
Uncertainty propagation & Report calibration, replicate, power, area and baseline contributions separately; generate confidence intervals by bootstrap or a stated probabilistic model. \\
Native files & Provide instrument-native files plus lossless comma-separated-value exports with timestamps, units, calibration identifiers and sample identifiers. \\
Operator-ready table & Wavelength, absorbed power, product rate, partial current, uncertainty, device geometry, polarization, incidence angle, potential, pH and temperature in one machine-readable table. \\
Mechanism assignment & Report direct, hot-carrier and photothermal contributions as constrained intervals or model comparison, not as unique fractions unless controls identify them. \\
Ledger inputs & Provide light and dark total current, full-cell voltage, product composition/flow, separation target and thermal boundary conditions needed for net-work accounting. \\
Reproducibility record & Analysis scripts, environment specification, random seeds, preregistered exclusions and a data-integrity manifest. \\
\bottomrule
\end{tabularx}
\end{table}

\clearpage
\section*{Supplementary References}